\documentclass[aps,nofootinbib,amsmath,prd,onecolumn,notitlepage,showpacs,superscriptaddress,11pt]{revtex4-1}

\allowdisplaybreaks
\usepackage[american]{babel}
\usepackage{amsfonts}
\usepackage{amsmath}
\usepackage{amssymb}
\usepackage{epsfig}
\usepackage{latexsym}
\usepackage{paralist}
\usepackage{fancyhdr}
\usepackage{graphicx}
\usepackage{etex}
\usepackage{braket}
\usepackage{float}
\usepackage{slashed}
\usepackage{xcolor}
\usepackage{soul}
\usepackage{dcolumn} 
\usepackage{bm}
\usepackage{nicefrac}
\usepackage{tipa}
\usepackage{dsfont}
\usepackage{xcolor}
\usepackage{slashed}
\usepackage{mathbbol}
\usepackage{hyperref}
\newcommand{\hnabla}{\hat{\nabla}}

\begin{document}

\title{Hidden gauge invariances of torsion theories:\\ closed algebras and absence of ghosts}

\author{Dario Sauro}
\email{dario.sauro@phd.unipi.it}
\affiliation{
Universit\`a di Pisa, Largo Bruno Pontecorvo 3, 56127 Pisa, Italy}


\begin{abstract}
%
We study the possible affine gauge transformations of the torsion tensor that make up Lie algebras. We find two such non-trivial structures, in which the gauge parameters are a $2$-form and a scalar. The first one gives rise to a non-abelian Lie algebra that is isomorphic to the Lorentz algebra, and which commutes with the latter. By linearizing this new gauge transformation we single out the gauge-invariant field variables on a flat background. Then, taking into account the most general power-counting renormalizable action of the torsion and imposing gauge invariance, we are able to find an invariant action that only has two free parameters. The torsion field equations of the resulting action are compatible with General Relativity in the torsionfree limit. Then, employing the spin-parity decompositions and the path integral method, we show that the theory, supplemented by the higher-derivative and Einstein-Hilbert actions, is free from kinematical ghosts, while it has a tachyonic scalar mode. Eventually, we comment on the technical difficulties that one will encounter when calculating the divergent part of the $1$-loop effective action.
\end{abstract}

\pacs{}
\maketitle

\renewcommand{\thefootnote}{\arabic{footnote}}
\setcounter{footnote}{0}

\section{Introduction}\label{sect:intro}

The theory of General Relativity (GR) \cite{Einstein:1916vd} is well-suited for the description of gravitational phenomena at low-energy scales, but it lacks consistency as we try to push the energy of a given process towards the Planck scale \cite{Goroff:1985th,Donoghue:1994dn}. For this reason throughout the years many possible high-energy completions of General Relativity have been proposed, like string theory \cite{Becker:2006dvp}, loop quantum gravity \cite{Ashtekar:1986yd}, asymptotic safety \cite{Wetterich:1992yh,Gies:2006wv,Percacci:2017fkn} and Lee-Wick Quantum Field Theory \cite{Anselmi:2017lia,Anselmi:2017yux}.

Another candidate UltraViolet (UV) completion of GR is provided by Metric-Affine theories of Gravity (MAGs), which retain the geometric spirit of the original formulation, while relaxing the metricity and torsionlessness assumptions. More concretely, the spacetime affine-connection $\hat{\Gamma}^\rho{}_{\nu\mu}$ is treated on an equal footing w.r.t.\ the metric $g_{\mu\nu}$ \cite{Palatini:1919ffw,Kibble:1961ba,Sciama:1964wt,Hehl:1976kj,Hehl:1994ue,Gronwald:1995em,Shapiro:2001rz,Hammond:2002rm}, thus the dynamical degrees of freedom generically differ from the sole graviton of GR. The GR description is usually recovered in the low-energy limit, when the torsion-free $T(\hat{\Gamma})=0$ and metric condition $\hat{\nabla} g = 0$ are retrieved as a dynamical consequence of the field equations \cite{Percacci:2023rbo}.  Furthermore, the metric-affine approach may also be extended to Yang-Mills theories, by assuming that the Hermitian form is spacetime-dependent and not covariantly constant \cite{Wachowski:2024nfp}.

Recently, metric-affine theories have received a renewed interest, largely due to the study of the so-called teleparallel equivalent theories \cite{BeltranJimenez:2019esp}. The latter are classically dynamically equivalent to GR, due to the fact that their Lagrange densities are equal to the Einstein-Hilbert one up to boundary terms. Here the two kinematical \emph{assumptions} that are presumed in Riemannian geometry (i.e., $T(\hat{\Gamma})=0$ and $\hat{\nabla} g = 0$) are replaced by the hypothesis that the full curvature tensor vanishes. The vanishing of this tensor brings about the notion of absolute teleparallelism, whence the name of these theories. Notice that this type of classical equivalence was already observed long ago in \cite{VanNieuwenhuizen:1981ae} for metric theories of torsion. On the other hand, recently the assumptions upon which the results of teleparallel equivalent theories rest have been criticized in \cite{Golovnev:2024lku}. As a matter of fact, the vanishing of the full curvature tensor has to be implemented through a Lagrange multiplier, and it does not descend from a symmetry principle. Thus, since the guiding principle of the present paper is to derive the dynamics from gauge principles and we want to avoid kinematical hypothesis, we shall discard teleparallel equivalent theories henceforth.

Whenever the metric tensor is non-degenerate, one is allowed to decompose the affine-connection into the sum of the Levi-Civita part plus a distortion. The latter receives two contributions, one coming from the contortion that is linear in the torsion tensor, and the other from the disformation which depends linearly on the non-metricity. From a gauge-theoretical viewpoint the contortion is associated with the Lorentz subgroup of the General Linear group $GL(d)$, whilst the disformation is linked to its orthogonal complement \cite{Gronwald:1995em}. Thus, metric-affine theories in which the non-metricity vanishes are those in which the local gauge group is the Lorentz one.

Even though metric-affine theories have been studied for a long time, little is known about their general features as Quantum Field Theories (QFTs) \cite{Melichev:2023lwj,Melichev:2024hih} (see also \cite{Gies:2022ikv,Balachandran:2024ktf} for the special case of Palatini gravity). Indeed, only recently their flat-space kinetic terms have been derived \cite{Baldazzi:2021kaf}, and this result somehow suggests that ghosts (i.e., negative norm states) and tachyons (states with imaginary masses) are widely present in these models (see also \cite{Marzo:2021iok,Barker:2022kdk,Barker:2024ydb,Barker:2024dhb,Barker:2024juc,Marzo:2024pyn}). Moreover, especially due to the complicated tensor structure of these fields, no known general statements regarding the (non-)renormalizability of these theories have been put forward. Nonetheless, in spite of the difficulties in working out systematic quantum predictions, metric-affine theories have been employed extensively for phenomenological purposes (see, e.g., \cite{Cai:2015emx,Aoki:2020zqm,Bahamonde:2020fnq,Karananas:2021gco,Iosifidis:2021bad,Bahamonde:2022kwg,Rigouzzo:2023sbb,Koivisto:2023epd,Mondal:2023cxx,Iosifidis:2024bsq,Karananas:2024xja,Koivisto:2024asr}).

As a matter of fact, using Feynman diagrams to perform quantum computations in MAGs is a formidably difficult task, for there are order a hundred possible interaction terms \cite{Baldazzi:2021kaf}. To our knowledge, the only attempt to use Feynman diagrams to compute $1$-loop corrections in MAGs is \cite{Marzo:2024pyn}, and these results suggest the need of having interactions constrained by gauge symmetries. On the other hand, also the covariant heat kernel technique \cite{DeWitt:1964mxt,Obukhov:1983mm,Vassilevich:2003xt} can hardly be used, because the differential operators that arise in generic Lagrangians involve non-minimal terms \cite{Martini:2023apm}. Indeed, the heat kernel coefficients for such operators are known only when one integrates out vectors, and they do not account for derivative interactions \cite{Barvinsky:1985an,Barvinsky:2021ijq}. Besides, the complete counterterm Lagrangian also comprises contributions coming from the higher-derivative operator acting on the metric \cite{Stelle:1977ry,Stelle:1976gc,Buoninfante:2023ryt}, as well as the third-order off-diagonal mixing between metric and torsion (or non-metricity) quantum fluctuations. Once again, the general form of the Seeley-DeWitt coefficients of the heat kernel due to these mixing terms is not known for minimal operators, and the situation is far more complicated when we take into account non-minimal operators. Thus, the relevance of the model that we propose in this paper is twofold: on the one hand it drastically reduces the number of free parameters, and on the other it allows selecting a minimal gauge for the torsion fluctuations, even though non-minimal terms still arise in the ghost action.   

During the years many models of torsionful theories of gravity have been proposed \cite{Yang:1974kj,Hehl:1978yt,Fairchild:1977wi,Mansouri:1976df,Tseytlin:1981nu}, some of them motivated by the absence of ghosts and tachyons when the theory is expanded around flat space \cite{Neville:1979rb,Sezgin:1979zf,Sezgin:1981xs}. Nevertheless, since there is no symmetry requirement that is employed to derive these models, they are not protected against the appearance of counterterms that are not already present in the bare Lagrangian. In this paper we try and fill this gap in the literature, by studying a class of infinitesimal affine transformations of the torsion that yield a closed algebra. In spite of this, in the metric-affine landscape the sole introduction of vector potential for Weyl symmetry does start from symmetry-based arguments and results in a fairly simple theory \cite{Ghilencea:2018thl,Ghilencea:2019jux}. However, its relationship with MAGs is often discarded.

From the field-theoretical viewpoint, the torsion is a vector-valued $2$-form, which can be split into its $GL(d)$-irreducible components. These are given by a $3$-form and a hook-antisymmetric rank-$3$ tensor. In the higher-spin literature the actions invariant under local gauge transformations for such higher-rank tensor fields have been derived \cite{Fronsdal:1978rb,deWit:1979sib,Didenko:2014dwa,Ponomarev:2022vjb}, and the relevant results for the irreducible components of the torsion have been investigated \cite{Labastida:1986gy,Labastida:1987kw,Metsaev:1997nj}. Moreover, the interplay between the St\"uckelberg mechanism and generalized symmetries has been undercovered recently in \cite{Chatzistavrakidis:2024dkw}. In this paper the St\"uckelberg mechanism for a hook antisymmetric tensor is discussed for the first time, though this tensor is not interpreted as an irreducible part of the spacetime torsion. Nevertheless, these results rest upon the simplifying assumption of living in a background maximally symmetric spacetime. Thus, even though heuristically our quest is similar to that of higher-spin theories, the technique and the scope of our investigation are different. Indeed, our background Riemannian geometry is completely general, and we do not assume any special or simplifying properties of the torsion. On top of this, we derive an invariant Lagrangian by requiring invariance under a non-abelian Lie-algebra. Only afterwards we specialize on a flat-space background, analyzing the stability of the model.

The infinitesimal transformations that we take into account act on the torsion and are the most general ones that are linear in the parameters and at most linear in the torsion itself. Thus, since the torsion counts as one derivative, the affine part of these transformations necessarily contains a single covariant derivative of the parameters, and the latter are forced to be even-rank tensors. We first consider the transformations that are $GL(d)$-irreducible, and in this case we find one solution parameterized by a scalar. By further allowing the mixing of the irreducible parts of the torsion we obtain another solution which is parameterized by a $2$-form. Most notably, this second solution has only two free parameters in the torsion sector of the action, and this number increases to a total of four when the higher-derivative and Einstein-Hilbert terms of the purely Riemannian sector are included.

In contrast with what has been commonly done in the metric-affine literature \cite{Neville:1979rb,Sezgin:1979zf,Percacci:2020bzf,Baldazzi:2021kaf}, we will not rely on spin-projectors to analyze the flat-space physics of our model. Instead, we shall follow the route paved by Mazur and Mottola in \cite{Mazur:1990ak,Mottola:1995sj}, which employs the path integral formulation and the covariant decomposition into spin-parity eigenstates \cite{york1974covariant,Martini:2023rnv}. Indeed, already at the level of the Einstein-Hilbert action these two schemes provide different outcomes. In fact, while spin-projectors seem to suggest that there is a ghost instability in the scalar sector \cite{Sezgin:1979zf}, the path integral method shows that the full functional determinant over the $0^+$ mode is just a constant. This is due to the cancellation that occurs when we take into account the non-trivial Jacobian in the path integral measure. We mention that the avoidance of propagation of such unphysical states can be achieved by a modification of the usual spin-projector formalism, see \cite{Marzo:2021esg}.   

Even though we will adhere with the method of the stability analysis of spin-parity eigenstates presented in \cite{Mazur:1990ak,Mottola:1995sj}, our presentation will nevertheless differ in one substantial point. Namely, we will be working in Euclidean signature from the onset, for this choice is actually mandatory whenever one employs a covariant spin-parity decomposition. Indeed, if the signature is Minkowskian the notion of orthogonality is too weak to carry over the standard proof of the existence and uniqueness of any such decomposition \cite{york1974covariant}. As a matter of fact, York's decomposition is commonly performed on the $3$-dimensional space-like hypersurfaces \emph{after} the $3+1$ decomposition of metric fluctuations. Eventually, only by breaking the symmetry between time and space coordinates we can unambiguously identify the propagating degrees of freedom, which are given by those whose Hessian includes the d'Alembert operator. For example, only the space-like Laplacian appears in the Hessian of the spin-$0$ gauge-invariant mode of the Einstein-Hilbert action, therefore this mode can give rise to hairs and not to particles.

Our definition of the path integral measure is based on the normalization of the Gaussian integral \cite{Mazur:1990ak}. Then, by performing a change of variable to the spin-parity eigenstates we obtain a non-trivial Jacobian. Next, by assuming the absence of the multiplicative anomaly we combine the functional determinants due to the measure and with the dynamical ones coming from the action functional. The outcome is a product of functional determinants over second-order differential operators evaluated on transverse and traceless tensors. Thence, our condition for the absence of ghost instabilities is that these differential operators must be positive semi-definite. We anticipate that there is a region in parameter space that is devoid of such instabilities. On the other hand, tachyonic modes are identified as mass terms with the wrong sign, and we do find one such mode in the $0^+$ spin-parity sector.

The paper is organized as follows. In the first part \ref{sect:intro-to-mags} we introduce the necessary definitions of torsionful theories of gravity. In Section \ref{sect:non-lin} we build the ansatze of the possible transformations, we and discuss the properties of the two transformations that yield closed algebras, analyzing the non-abelian one in detail. In particular, we show that the non-abelian algebra is isomorphic to the Lorentz one, that it commutes with the latter and that we have the usual semi-direct structure when diffeomorphisms are considered. In Sect.\ \ref{sect:gauge-inv-dof} we review the covariant spin-parity decompositions of the torsion and metric fluctuations, singling out the gauge-invariant field variables and defining the path integral measures on a flat background. In Sect.\ \ref{sect:inv-act} we derive the invariant action that we have previously mentioned and compute the part of the field equations which is necessary for drawing conclusions regarding the compatibility with GR. We also introduce higher-derivative metric terms in the action, as these would be generated radiatively. In Sect.\ \ref{sect:particle-content} we perform the spin-parity decomposition of the linearized action on a flat-background, we build the partition function and we show the absence of ghost instabilities, commenting on the evaluation of $1$-loop quantum fluctuations on a generic background in \ref{subsect:quantum-remarks}. Eventually, in \ref{sect:conclusions} we conclude with some general remarks and outlooks. The two appendices contain the details of the closure of the non-abelian algebra \ref{sect:app-closed-algebra} and of the derivation of the action \ref{sect:app-derivation-action}, which were not included in the main body of the article due to their length.

\section{A short introduction to MAGs}\label{sect:intro-to-mags}

The most general geometry that we encounter in metric affine-theories is described in terms of the torsion, the non-metricity and the spacetime curvature. Only the latter is non-zero in General Relativity, where the physical consequences of gravitational phenomena are described solely in terms of the metric tensor. On the other hand, if we allow the affine-connection to be a dynamical field, we generically end up with a metric-affine geometry.

Given an affine-connection $\hat{\Gamma}^\rho{}_{\nu\mu}$, we write down the covariant derivative of a contra-variant vector as
\begin{equation}
 \hat{\nabla}_\mu v^\rho \equiv \partial_\mu v^\rho + \hat{\Gamma}^\rho{}_{\nu\mu} v^\nu \, .
\end{equation}
Then, the three tensors that encode gravitational effects in metric-affine theories are defined by
\begin{subequations}
 \begin{align}\label{eq:def-torsion}
  & T^\rho{}_{\mu\nu} \equiv \Gamma^\rho{}_{\nu\mu} - \Gamma^\rho{}_{\mu\nu} \, ;\\
  & Q_{\mu\alpha\beta} = - \hat{\nabla}_\mu g_{\alpha\beta} \, ;\\\label{eq:def-curvature}
  & \hat{R}^\rho{}_{\lambda\mu\nu} = 2 \left( \partial_{[\mu} \hat{\Gamma}^\rho{}_{|\lambda|\nu]} + \hat{\Gamma}^\rho{}_{\alpha[\mu} \hat{\Gamma}^\alpha{}_{|\lambda|\nu]} \right) \, .
 \end{align}
\end{subequations}
These tensors are known as the torsion, the non-metricity and the curvature, respectively.

If the metric is non-singular, we can define the Levi-Civita connection $\Gamma^\rho{}_{\nu\mu} = g^{\lambda\rho} \partial_{(\mu} g_{\nu) \lambda} - \frac{1}{2} \partial^\rho g_{\mu\nu}$, and split the full affine connection as
\begin{equation}\label{eq:split-aff-conn}
 \hat{\Gamma}^\rho{}_{\nu\mu} = \Gamma^\rho{}_{\nu\mu} + K^\rho{}_{\nu\mu} + N^\rho{}_{\nu\mu} \, ,
\end{equation}
where $K^\rho{}_{\nu\mu}$ is the contortion tensor and $N^\rho{}_{\nu\mu}$ is the disformation. The latter are written in terms of the torsion and non-metricity, respectively
\begin{subequations}
 \begin{align}\label{eq:def-contortion}
  K^\rho{}_{\nu\mu} & = \frac{1}{2} \left( T_\nu{}^\rho{}_\mu + T_\mu{}^\rho{}_\nu - T^\rho{}_{\nu\mu} \right) \, ;\\
  N^\rho{}_{\nu\mu} & = \frac{1}{2} \left( Q_\mu{}^\rho{}_\nu + Q_\nu{}^\rho{}_\mu  - Q^\rho{}_{\mu\nu} \right) \, .
 \end{align}
\end{subequations}
Notice that the contortion is antisymmetric in the first two-indices, thus it takes values in the orthogonal sub-algebra of $GL(4)$. On the other hand, the non-metricity is symmetric in its lowers indices, thence it does not contribute to the torsion. 
In what follows we shall be interested in torsionful theories, therefore we will set the non-metricity to zero from the onset.

Historically, MAGs have been formulated as the gauge theory of the local affine group $Gl(4)\rtimes T_4$, that is attached to every spacetime point, where the second factor stands for translations. Both $Gl(4)$ and $T_4$ are endowed with their own gauge potentials, that allow us to covariantly differentiate tensors that bear internal $Gl(4)$ indices (labeled by the first letters of the Latin alphabet). The gauge potential of $Gl(4)$ reduces to the spin-connection as we let $Gl(4) \rightarrow SO(4)$, whereas this role is played by the soldering form $\theta^a$ when it comes to translations, and its gauge covariant part is the co-frame $e^a$. Thus, when only the local Lorentz group is retained, the torsion can alternatively be defined as the exterior covariant derivative of the co-frame, i.e.,
\begin{align}
 T^a = & \, d e^a + \omega^a{}_b \wedge e^b   \\\nonumber
 = & \, e^a{}_\rho T^\rho{}_{\mu\nu} dx^\mu \wedge dx^\nu \, .
\end{align}
On the other hand, in the most general metric-affine case the local connection $A^{a}{}_{b\mu}$ takes values in the $\mathfrak{gl}(4)$ algebra. The latter can be written as the direct sum of the special linear algebra and dilatations, i.e.,
\begin{equation}
	\mathfrak{gl}(4) = \mathfrak{sl}(4) \oplus \mathfrak{D}(1) \, . 
\end{equation}
Furthermore, the special orthogonal group $SO(4)$ arises as the maximal compact subgroup of the special linear group. In terms of the generators $J^{ab}$ of the $\mathfrak{gl}(4)$ algebra these splittings can be understood in the following manner. The trace part of $J^{ab}$ generates the one-dimensional dilatation abelian algebra, i.e., $J^a{}_a \in \mathfrak{D}(1)$. In contrast, by taking the antisymmetric part of $J^{ab}$ we find all the generators of the rotation algebra, i.e., $J^{[ab]}\in\mathfrak{so}(4)$. Finally, the remaining generators, that is, those which are symmetric and traceless, generates the constant-volume shears. It is important to remark that the commutator of two isochor shear generators yields a rotation, which is why these members of the $\mathfrak{sl}(4)$ algebra cannot provide any nontrivial sub-algebra.

Let us now apply this geometric framework to metric-affine theories; the first tensor that we need to consider is the local metric $g_{ab}$, which is not covariantly constant whenever the local gauge group is not the special orthogonal one. Indeed, its covariant derivative is given by
\begin{equation}
	\hnabla_\mu g_{ab} = \partial_\mu g_{ab} - A^c{}_{a\mu} g_{cb} - A^c{}_{b\mu} g_{ac} = - 2 A_{(ab)\mu} \, .
\end{equation}
Local and holonomic indices are converted through the contraction of the co-frame $e^a{}_\mu$ or its inverse $e^\mu{}_a$, whose covariant derivative is postulated to vanish as an integrability condition which glues together the local and tangent spaces, i.e,
\begin{equation}
	\hnabla_\mu e^a{}_\nu = 0 \, .
\end{equation}
This equation allows us to connect the non-metricity tensor to the symmetric part of the local connection $A^a{}_{b\mu}$ as
\begin{equation}
	\begin{split}
		Q_{\mu\nu\rho} = & - \hnabla_\mu g_{\nu\rho} = - e^a{}_\nu e^b{}_\rho \hnabla_\mu g_{ab} \\
		= & 2 e^a{}_\rho e^b{}_\nu A_{(ab)\mu} \, .
	\end{split}
\end{equation}
Therefore, the presence of non-metricity intrinsically implies that the underlying local gauge structure has been augmented with respect to the Riemannian case, also including dilations and isochor shears. Since the latter transformations do not preserve the line element of any point-particle, the introduction of non-metricity profoundly changes the geometric assumptions of a putative modified gravity theory. For this reason, we shall discard the non-metricity in this paper, considering only the so-called antisymmetric MAGs, whose starting hypotheses differ only slightly from those of the Einstein theory.

In analogy with Eq.\ \eqref{eq:split-aff-conn}, also the spin-connection can be written as the sum of the Levi-Civita contribution (which is torsionfree), and the torsionful part, namely
\begin{equation}\label{eq:split-spin-conn}
 \omega^a{}_b = \mathring{\omega}^a{}_b + \Omega^a{}_b \, .
\end{equation}
The first term on the right-hand-side is written in terms of the co-frame, its derivative and the inverse co-frame $E^\mu{}_a$ as
\begin{equation}\label{eq:lc-spin-conn}
\mathring{\omega}^a{}_{b\mu} = \frac{1}{2} \left( E^{\nu}{}_b (\partial_\nu e^a{}_\mu - \partial_\mu e^a{}_\nu) + E^{\nu a} (\partial_\mu e_{b\nu} - \partial_\nu e_{b\mu}) + E^{\rho a} E^\nu{}_b e_{c\mu} (\partial_\nu e^c{}_\rho - \partial_\rho e^c{}_\nu ) \right) \, ,
\end{equation}
while $\Omega^a{}_b$ is isomorphic to the contortion due to the co-frame postulate
\begin{equation}
 \hat{\nabla}_\mu e^a{}_\nu = 0 \, .
\end{equation} 
This last equation simply states that only one among the spin-connection and the affine-connection can be dynamical. Choosing between the two of them is arbitrary, since we can perform calculations employing either coordinate or local indices.

Having settled the basic geometric definitions that we shall need in the rest of the paper, we proceed to the analysis of the consistent affine transformations of the torsion.

\section{Non-linear transformations}\label{sect:non-lin}

This section is devoted to the study of the possible infinitesimal affine transformations of the torsion tensor. As we have explained in the Introduction \eqref{sect:intro}, we shall focus only on those variations that yield closed algebras.

\subsection{The ansatze}\label{subsect:ansatze}

First of all, we want to consider infinitesimal affine transformations, i.e., those that are of the following schematic form
\begin{equation}\label{eq:transf-ansatz}
 T^\rho{}_{\mu\nu} \rightarrow T^\rho{}_{\mu\nu} + (\nabla \Phi)^\rho{}_{\mu\nu} + (\Phi T)^\rho{}_{\mu\nu} \, ,
\end{equation}
where $\Phi$ are tensor fields that parameterize the possible infinitesimal transformations. To attack the problem, we will first consider the algebraic irreducible terms of the form $(\nabla \Phi)^\rho{}_{\mu\nu}$. The first constraint that comes from a trivial tensor analysis is that $\Phi$ must have an even number of indices. Combining this fact with the known spin-parity content of the torsion implies that $\Phi$ can be either a scalar, a symmetric $2$-tensor or two antisymmetric $2$-tensors \cite{Aoki:2019snr,Baldazzi:2021kaf,Martini:2023rnv}. Here we assume that the completely antisymmetric Levi-Civita tensor is not involved in the ansatz \eqref{eq:transf-ansatz}. Thus, the tensors that paremeterize the viable transformations are necessarily parity even, and the Levi-Civita tensor cannot enter the structure constants of any algebra.

We start from the the symmetric tensor $S_{\mu\nu}$: since the affine part of the transformation is necessarily hook antisymmetric, we impose such a symmetry on the homogeneous part of the transformation too. Thus, the first infinitesimal transformation of the torsion that we study is given by
\begin{align}\label{eq:ansatz-deltaS}
 \delta^1{}_S T^\rho{}_{\mu\nu} = \, & \nabla_\mu S^\rho{}_\nu - \nabla_\nu S^\rho{}_\mu + 2  a_1 \left( T^\rho{}_{[\mu}{}^\lambda S_{\nu]\lambda} + T_{[\nu\mu]}{}^\lambda S^\rho{}_\lambda \right)\\\nonumber
& + a_2 \left( 2 T^\lambda{}_{\mu\nu} S^\rho{}_\lambda + T^{\lambda\rho}_\nu S_{\mu\lambda} - T^{\lambda\rho}_\mu S_{\nu\lambda} \right) + a_3 \left( T^\lambda{}_{\mu\lambda} S^\rho{}_\nu - T^\lambda{}_{\nu\lambda} S^\rho{}_\mu \right) \, .
\end{align}
Notice that there are three independent hook antisymmetric contractions of the $S_{\mu\nu}$ tensor to the torsion, and the last one involves the torsion vector. We remain in the hook anti-symmetric sector, and write down the second infinitesimal transformation. In this case the parameter is a rank-$2$ tensor $A_{\mu\nu}$, and we have
\begin{align}\label{eq:ansatz-deltaA}
\delta^2{}_A T^\rho{}_{\mu\nu} = \, & 2 \nabla^\rho A_{\mu\nu} + \nabla_\mu A^\rho{}_\nu - \nabla_\nu A^\rho{}_\mu + c_1 \left( 2 T_{[\mu\nu]}{}^\lambda A^\rho{}_\lambda + 2 T^\rho{}_{[\nu}{}^\lambda A_{\mu]\lambda} \right)\\\nonumber
& + c_2 \left( - 2 T^\lambda{}_{\mu\nu} A^\rho{}_\lambda + 2 T^{\lambda\rho}{}_{[\nu} A_{\mu]\lambda} \right) + c_3 \left( T^\lambda{}_{\mu\lambda} A^\rho{}_\nu - T^\lambda{}_{\nu\lambda} A^\rho{}_\mu \right) \, .
\end{align}
As in the previous case, there are three independent contributions in the former expression, with one involving the torsion vector.

Now we turn to the completely antisymmetric transformation, which is parameterized by a $2$-form $\Pi_{\mu\nu}$. The most general ansatz is
\begin{align}\label{eq:ansatz-deltaPi}
\delta^3{}_\Pi T^\rho{}_{\mu\nu} = \, & \nabla^\rho \Pi_{\mu\nu} + \nabla_\mu \Pi_\nu{}^\rho + \nabla_\nu \Pi^\rho{}_\mu + 2 b_1 \left( T^\rho{}_{[\nu}{}^\lambda \Pi_{\mu]\lambda} + T_{[\mu}{}^{\rho\lambda} \Pi_{\nu]\lambda} + T_{[\nu\mu]}{}^\lambda \Pi^\rho{}_\lambda \right)\\\nonumber
& + b_2 \left( T^\lambda{}_\mu{}^\rho \Pi_{\lambda\nu} + T^\lambda{}_{\nu\mu} \Pi_\lambda{}^\rho + T^{\lambda\rho}{}_\nu \Pi_{\lambda\mu} \right) + b_3 \left( T^{\lambda\rho}{}_\lambda \Pi_{\mu\nu} + T^{\lambda}{}_{\mu\lambda} \Pi_\nu{}^\rho + T^{\lambda}{}_{\nu\lambda} \Pi^\rho{}_{\mu} \right) \, .
\end{align}
Lastly, we take the parameter to be a scalar, and in this case the transformation is necessarily hook-antisymmetric. This transformation can be viewed as a special case o Eq.\ \eqref{eq:ansatz-deltaS}. Accordingly, we have one less independent contraction with respect to the previous cases
\begin{align}\label{eq:ansatz-deltaphi}
 \delta^4{}_\phi T^\rho{}_{\mu\nu} = \, & \delta^\rho{}_\nu \partial_\mu \phi - \delta^\rho{}_\mu \partial_\nu \phi + z_1 \left( T^\rho{}_{\mu\nu} + T_\mu{}^\rho{}_\nu - T_\nu{}^\rho{}_\mu \right) \phi \\\nonumber
 & + z_2 \left( \delta^\rho{}_\nu T^\alpha{}_{\mu\alpha} - \delta^\rho{}_\mu T^\alpha{}_{\nu\alpha} \right) \phi \, .
\end{align}
Let us observe that this last transformation is a generalization of the affine transformation of the torsion vector proposed in \cite{Shapiro:2001rz}. Indeed, when $z_1=0$ only the vector irreducible representation is affected. Moreover, this special case can also be understood as an affine Weyl transformation of the torsion when $z_2=\frac{1}{d-1}$ \cite{Sauro:2022chz,Sauro:2022hoh,Zanusso:2023vkn,Paci:2024ohq}. 

\subsection{Closed algebras}\label{subsect:closure}

After having presented the ansatze of the possible affine transformations of the torsion, we take into account their algebraic structure. Since we want to inspect closed algebras, we only evaluate the commutator of a two transformations of the same type.
A given algebra closes when the commutator of two transformations can be expressed as a single transformation, written in terms of the commutator of the parameters. For irreducible rank-two tensors such a commutator is unique, and we fix the convention of this sign as
\begin{equation}
\left[ A_1, A_2 \right]_{\mu\nu} \equiv A_1{}_\mu{}^\lambda A_2{}_{\nu\lambda} - A_2{}_\mu{}^\lambda A_1{}_{\nu\lambda} \, .
\end{equation}
It is easily seen that the first class of transformations, which is parameterized by $S_{\mu\nu}$, cannot close unless it is abelian. On the other hand, when it comes to Eq.s \eqref{eq:ansatz-deltaA} and \eqref{eq:ansatz-deltaPi} we need to explicitly do the computation. Due to the length and tediousness of this task we employed some of the free \texttt{xAct} packages for \texttt{Mathematica} \cite{Martin-Garcia:xAct,Martin-Garcia:xTensor,Martin-Garcia:2008ysv,Nutma:2013zea}, and we shall not present anything but the final results here. What we find is that neither the second nor the third transformations yield closed algebras. Thus, there are no non-abelian algebraic structures generated by $GL(d)$ irreducible transformations.

The next logical step is to relax the hypothesis of irreducibility, and to take into account the linear combination of Eq.s\ \eqref{eq:ansatz-deltaA} and \eqref{eq:ansatz-deltaPi}, with $\Pi_{\mu\nu}$ replaced by $\theta A_{\mu\nu}$. Interestingly, we find that for $\theta=\frac{1}{4}$ the transformation yields a closed algebra. The explicit form of the former equation reads
\begin{align}\label{eq:deltaG}
 \delta^G{}_A T^\rho{}_{\mu\nu} = & \, 3 \nabla^\rho A_{\mu\nu} + \nabla_\mu A^\rho{}_\nu - \nabla_\nu A^\rho{}_\mu\\\nonumber
 & + \zeta^2 \left[ 2 T_{[\mu\nu]}{}^\lambda A^\rho{}_\lambda + \frac{2}{3} T_{[\mu}{}^{\rho\lambda} A_{\nu]\lambda} + \frac{2}{3} T^\lambda{}_{\mu\nu} A^\rho{}_\lambda + \frac{4}{3} T^\lambda{}_{[\mu}{}^\rho A_{\nu]\lambda} - \frac{10}{3} T^\rho{}_{[\mu}{}^\lambda A_{\nu]\lambda} \right] \, ,
\end{align}
where only one free parameter $\zeta$ is left undetermined. The detailed derivation of the closure of the algebra is relegated to the Appendix 	\ref{sect:app-closed-algebra}. Since this constant multiplies every term that is linear in the torsion, we may be tempted to draw an analogy with the Yang-Mills coupling constant. Moreover, this parallelism is strengthened by the fact that the commutator of two transformations takes the following form 
\begin{equation}\label{eq:algebra}
 \left[ \delta^G{}_{A_1} , \delta^G{}_{A_2} \right] = \frac{4}{3} \zeta^2 \, \delta^G{}_{[A_1,A_2]} \, .
\end{equation}

We have also analyzed the fourth affine transformation, finding that when $z_2 = - \frac{3z_1}{d-1}$ the transformation gives rise to an abelian algebra. Later in this section we shall come back to the properties of this transformation.

\subsection{Full gauge algebra of the theory}

By imposing new infinitesimal symmetries in a given system, we automatically have new Noether identities. We are now interested in the intertwining between the latter and the Lorentz Noether identities. In order to study it, we first have to derive the induced transformations of the co-frame and spin-connection that make up the torsion itself. This is because the Lorentz Noether identities cannot be derived automatically in terms of the Lorentz-invariant field variables $g_{\mu\nu}$ and $T^\rho{}_{\mu\nu}$.

At this point one may wonder why we have not started our analysis of the closed algebraic structures of torsionful theories directly from the co-frame and spin-connection. The reason for this choice is purely a matter of convenience, since dealing with coordinate-based tensors simplifies the calculations and renders them more easily implementable on a software such as \texttt{Mathematica}. More importantly, the validity of such an alternative formulation rests on our capability of unambiguously retrieving the form of the most general affine transformation of the co-frame and spin-connection parametrized by a two-form. Indeed, the only source of confusion may arise from the introduction of an arbitrary parameter $\gamma$ which tunes the transformation of the co-frame. Nevertheless, the physics is insensitive to this parameter, since both the invariant Lagrangian that we will derive in section \ref{sect:inv-act} and the Noether identities associated to the symmetry transformation do not depend on $\gamma$.

We attack the problem of finding the induced transformations of $e^a{}_\mu$ and $\omega^{ab}{}_\mu$ by splitting the spin-connection as in Eq.\ \eqref{eq:split-spin-conn} and writing its torsionful part as a function of the contortion and co-frame. Therefore, by employing Eq.s \eqref{eq:def-contortion} and \eqref{eq:deltaG} we first derive the variation of the contortion, that reads
\begin{align}\label{eq:deltaG-contortion}
K^\alpha{}_{\beta\mu} \rightarrow & K^\alpha{}_{\beta\mu} + \frac{5}{2} \nabla_\mu A^\alpha{}_\beta + \frac{3}{2} \nabla_\beta A^\alpha{}_\mu - \frac{3}{2} \nabla^\alpha A_{\beta\mu}\\\nonumber
& + \zeta^2 \left[ A^\lambda{}_\mu \left( 3 K^\alpha{}_{\beta\lambda} - \frac{5}{2} K^\alpha{}_{\lambda\beta} + \frac{5}{2} K_{\beta\lambda}{}^\alpha \right) + A^\lambda{}_\beta \left( K^\alpha{}_{\mu\lambda} - \frac{3}{2} K_{\lambda\mu}{}^\alpha - \frac{1}{6} K^\alpha{}_{\lambda\mu} \right) \right.\\\nonumber
& \left. \qquad\,\,\,\, - A^{\lambda\alpha} \left( K_{\beta\mu\lambda} - \frac{3}{2} K_{\lambda\mu\beta} - \frac{1}{6} K_{\beta\lambda\mu} \right) \right] \, .
\end{align}
Since $A_{\mu\nu}$ is a rank-$2$ tensor that counts as zero derivatives exactly like the the co-frame and its Weyl weight is $w(A_{\mu\nu})=2$, the variation of the co-frame is uniquely determined by tensor analysis, up to a constant
\begin{equation}\label{eq:deltaG-coframe}
e^a{}_\mu \rightarrow  \, e^a{}_\mu + \gamma A^a{}_\mu = e^a{}_\mu + \gamma A^\nu{}_\mu e^a{}_{\nu} \, .
\end{equation}
The variation of the co-frame is the only ansatz that we have to make in order to derive the gauge transformation of the field variables in the $\{e^a{}_\mu, \omega^a{}_{b\mu}\}$ basis. However, this ansatz is derived in a simple fashion by exploiting tensor analysis and a counting of derivatives. By inserting this transformation law into Eq.\ \eqref{eq:split-spin-conn} we obtain the transformation of the Levi-Civita part, that is
\begin{equation}
 \mathring{\omega}^{ab}{}_\mu \rightarrow \mathring{\omega}^{ab}{}_\mu - \gamma \nabla_\mu A^{ab} \, .
\end{equation}
Finally, using this result together with the variation of the contortion \eqref{eq:deltaG-contortion}, we derive the infinitesimal transformation law of the torsionful spin-connection
\begin{align}\nonumber\label{eq:deltaG-spin-connection}
 \omega^{ab}{}_\mu \rightarrow & \, \left( \frac{5}{2} -\gamma \right) \nabla_\mu A^{ab} - 3 \nabla^{[b} A^{a]}{}_\mu + \zeta^2 \left[ 3 A^c{}_\mu \Omega^{ab}{}_c - 5 A^c{}_\mu \Omega^{[a}{}_c{}^{b]} + 2 A^{c[b} \Omega^{a]}{}_{\mu c} - 3 A^{c[b} \Omega_{c\mu}{}^{a]} \right. \\
 & \, + \left. \left( \frac{2\gamma}{\zeta^2} + \frac{1}{3} \right) A^{[a}{}_c \Omega^{|c|b]}{}_\mu \right] \, .
\end{align}
Thus, the new symmetry transformation acts as a one-parameter family of local Lorentz rotations on the co-frame. However, the transformation of the spin-connection is more general than that of a connection $1$-form, since we not only have the exterior derivative of the rotation matrix, but we also encounter the rotor of the latter. In particular, in the limit of vanishing charge $\zeta\rightarrow0$ and by setting $\gamma=\tfrac{5}{2}$ the transformation is given by
\begin{equation}
	\delta \omega^a{}_{b\mu} \big|_{\zeta=0, \, \gamma=\tfrac{5}{2}} = \frac{3}{2} \nabla^a A_{b\mu} - \frac{3}{2} \nabla_b A^a{}_\mu + \frac{1}{6} A^a{}_c \Omega^c{}_{b\mu} - \frac{1}{6} A^c{}_b \Omega^a{}_{c\mu} \, ,
\end{equation}
i.e., it reduces to the sum of the rotor of the internal rotation matrix (times a numerical factor) plus a roto-dilation of the contortion.

At this point, we consider the commutator of the infinitesimal transformation \eqref{eq:deltaG} with the other two gauge transformations of the theory, i.e., local Lorentz ones and diffeomorphisms. We start from the first, and we choose to act on the co-frame to derive the commutator. An infinitesimal Lorentz transformation acts as
\begin{equation}
 \delta^L{}_\alpha e^a{}_\mu = \alpha^a{}_b \, e^b{}_\mu \, ,
\end{equation}
where $\alpha^{ab} = - \alpha^{ba}$. Thus, by noting that Eq.\ \eqref{eq:deltaG-coframe} acts on the co-frame field and not on its latin indices \emph{per se}, it is easy to see that the commutator vanishes
\begin{equation}\label{eq:deltaG-lorentz}
 [\delta^L{}_\alpha,\delta^G{}_A ] = 0 \, .
\end{equation}
We remark that we have chosen to derive the result acting on the co-frame, but it holds also when we apply these transformations on other fields, such as the spin-connection.

By putting together Eq.s\ \eqref{eq:algebra} and \eqref{eq:deltaG-lorentz} we deduce that the new gauge transformation $\delta^G$ may be interpreted as an holonomic copy of the orthogonal algebra. Indeed, $\delta^G$ is isomorphic to the latter, and they commute, i.e., they have the direct product structure $\delta^G \times \delta^L$.

Now we turn to the mixing with diffeomorphisms. For simplicity, also in this case we choose to act on the co-frame. Moreover, we employ the geometric language of differential $p$-forms, i.e., we omit the coordinate indices, since it simplifies the derivation of the result. The infinitesimal transformation of the co-frame under diffeomorphisms $\delta^E{}_\xi$ can be written employing the so-called Cartan's magic formula as
\begin{equation}
 \delta^E{}_\xi e^a = \left( d \circ \iota_\xi + \iota_\xi \circ d \right) e^a \, .
\end{equation}
Here $d$ is the exterior differential acting on differential $p$-forms, while $\iota_\xi$ is the contraction of a $p$-form with the contra-variant vector $\xi$, which we define by $\iota_\xi (dx^\mu \wedge dx^\nu) = \xi^\mu dx^\nu - \xi^\nu dx^\mu$. Thus, by employing the previous expression and Eq.\ \eqref{eq:deltaG-coframe}, we readily find that
\begin{equation}
\left[ \delta^G{}_{A} , \delta^{E}{}_{\xi} \right] e^a = \gamma \, \iota_\xi ( d A^a{}_b) \, e^b \, .
\end{equation}
Accordingly, we have that the resulting gauge structure is given by a	 semi-direct product, i.e.,
\begin{equation}
\left[ \delta^G{}_{A} , \delta^{E}{}_{\xi} \right] e^a{}_\mu = \delta^G{}_{\xi^\nu\partial_\nu A} \, .
\end{equation}
Finally, since this semi-direct structure holds also when we take into account the mixing of diffeomorphisms and Lorentz transformations, the full gauge algebra $\mathbb{g}$ of our theory is
\begin{equation}\label{eq:full-gauge-algebra}
 {\mathbb g} = \left(\delta^G \times \delta^L \right) \ltimes \delta^{E} \, .
\end{equation}

Having studied the algebraic structure of this gauge transformation of the torsion, we conclude this section focusing on the Noether identities that stem from invariance under \eqref{eq:deltaG}.

\subsection{Noether identities and coupling to matter fields}

In this subsection we briefly derive the Noether identity that is associated with our new local gauge transformation \eqref{eq:deltaG}. Since the theory already has two non-trivial Noether identities stemming from local Lorentz invariance and diffeomorphism invariance, we also discuss how these intertwine with our newly found gauge transformation \eqref{eq:deltaG}.

The current that is associated with the functional dependence of a matter action $S_m$ on the co-frame will be referred to as the non-symmetric energy-momentum tensor and is defined as
\begin{equation}\label{eq:non-sym-energy-mom}
T^\mu{}_a = - \frac{1}{\boldsymbol{e}} \frac{\delta S_m}{\delta e^a{}_\mu} \, ,
\end{equation}
where $\boldsymbol{e}$ is the determinant of the co-frame. On the other hand, the dependence on the spin-connection is parameterized by the spin-current, whose definition is
\begin{equation}\label{eq:spin-current}
\Sigma^\mu{}_{ab} = \frac{1}{\boldsymbol{e}} \frac{\delta S_m}{\delta \omega^{ab}{}_\mu} \, .
\end{equation}
These two currents have their counterparts when the theory does not involve fermions, and thence it can be written entirely in terms of the metric and torsion tensors. In such a case, we define the energy-momentum tensor as
\begin{equation}\label{eq:energy-mom}
\Theta^{\mu\nu} = - \frac{2}{\sqrt{-g}} \frac{\delta S_m}{\delta g_{\mu\nu}} \, .
\end{equation}
Similarly, we define the torsion-current as
\begin{equation}\label{eq:torsion-current}
Z_\lambda{}^{\mu\nu} = \frac{2}{\sqrt{-g}} \frac{\delta S_m}{\delta T^\lambda{}_{\mu\nu}} \, .
\end{equation}
The Noether identities generated by invariance under local Lorentz and diffeomorphism transformations are usually written in terms of the first two currents, and they read, respectively,
\begin{subequations}
 \begin{align}\label{eq:nother-lorentz}
  & T_{[\alpha\beta]} + (\hat{\nabla}_\mu + \tau_\mu) \Sigma^\mu{}_{\alpha\beta} \, ;\\\label{eq:nother-einstein}
  & (\hat{\nabla}_\mu + \tau_\mu) T^\mu{}_\nu = T^\mu{}_\rho T^\rho{}_{\nu\mu} - \Sigma^\mu{}_{\alpha\beta} \hat{R}^{\alpha\beta}{}_{\nu\mu} \, .
 \end{align}
\end{subequations}

Since the torsion and metric tensor depend on the co-frame and spin-connection, we can write the first two currents Eq.s \eqref{eq:non-sym-energy-mom} and \eqref{eq:spin-current} in terms of $\Theta^{\mu\nu}$ and $Z_\lambda{}^{\mu\nu}$. This just amounts to using the functional chain rule, obtaining
\begin{subequations}
 \begin{align}\label{eq:non-sym-energy-to-sym}
  & T^{\mu\nu} = \Theta^{(\mu\nu)} + (\hat{\nabla}_\alpha + \tau_\alpha) Z^{\nu\alpha\mu} \, ; \\\label{eq:spin-current-to-torsion-current}
  & \Sigma^\mu{}_{\rho\nu} = Z_{[\rho}{}^\mu{}_{\nu]} \, .
 \end{align}
\end{subequations}
It is then easy to see that the Lorentz Noether identity \eqref{eq:nother-lorentz} is identically satisfied when we substitute the expressions of the symmetric energy-momentum and torsion-current. This is the reason why Lorentz invariance cannot constrain the parameter space of any torsionful theory of gravity.

Thus, employing the definition of the torsion-current \eqref{eq:torsion-current} and the infinitesimal transformation rule of the torsion, we obtain the Noether identity that is associated with invariance under $\delta^G$, i.e.,
\begin{align}\label{eq:noetherG}
 & 3 \nabla_\lambda Z^{\lambda\mu\nu} + \nabla_\lambda Z^{\nu\mu\lambda} - \nabla_\lambda Z^{\mu\nu\lambda} + \zeta^2 \left[ \frac{10}{3} T^{\alpha[\nu|\beta|} Z_\alpha{}^{\mu]}{}_\beta + \frac{4}{3} T^{[\mu|\alpha\beta|} Z_\alpha{}^{\nu]}{}_\beta \right.\\\nonumber
 & \left.  + \frac{2}{3} T^{\alpha[\mu|\beta|} Z_\beta{}^{\nu]}{}_\alpha + \frac{2}{3} T^{[\mu}{}_{\alpha\beta} Z^{\nu]\alpha\beta} + 2 T^{\alpha[\nu|\beta|} Z^{\mu]}{}_{\alpha\beta} \right] = 0 \, .
\end{align}

The previous equation allows us to briefly discuss the invariant coupling of the torsion with matter and radiation fields. Indeed, $Z_\lambda{}^{\mu\nu}$ is the current associated to the torsion, and it stands for the first functional derivative of the matter action w.r.t. the torsion. It turns out that there is no invariant coupling of matter fields with spin $0$, $\frac{1}{2}$ and $1$ to the torsion. Thus, the torsion is decoupled from Standard Model fields, which can only be coupled to the Riemannian structure of the spacetime.

After having studied these general properties of the new affine transformation that we have found, in the next section we turn to some of the dynamical consequences that are obtained by imposing it.

\section{Gauge-invariant degrees of freedom}\label{sect:gauge-inv-dof}

In this section we shortly discuss the covariant spin-parity decomposition of the metric and torsion fluctuations, and we employ them to obtain the gauge-invariant degrees of freedom.

Any spin-parity decomposition enables us to split the perturbations into their transverse and longitudinal components. Then, by linearizing the gauge transformations of our theory we can single out the gauge-invariant fields by performing a spin-parity decomposition on the parameter of the gauge transformation. We analyze both diffeomorphism and $\delta^G$ as gauge transformations, reviewing the results of the analysis for the first invariance group to outline the general method. 

Afterwards, we review how these splittings into spin-parity eigenstates suggest a change of variable in the quantum functional measures. In Sect.\ \ref{subsect:particle-content} we will employ these results to show that taking into account the Jacobian of this change of variable effectively reduces the order of the differential operators that act on the longitudinal components.

\subsection{Spin-parity decomposition in the flat-space limit}

Now we aim at singling out the gauge-invariant degrees of freedom of the torsion and metric field and to inspect their propagation on top of a flat-space background. To this end, we shall make use of the flat-space covariant decomposition of the metric and torsion tensors \cite{york1974covariant,Martini:2023rnv}, accompanied by the Hodge decomposition of the gauge parameters. Since any spin-parity decomposition is well-posed only when the spacetime signature is positive definite \cite{york1974covariant}, we always work in Euclidean signature.

Given a metric fluctuation $h_{\mu\nu}$, the flat-space version of the York decomposition reads
\begin{align}\label{eq:york}
h_{\mu\nu} = h_{\mu\nu} + \partial_\mu \xi_\nu + \partial_\nu \xi_\mu + \partial_\mu \partial_\nu \sigma - \frac{1}{d} \delta_{\mu\nu} \partial^2 \sigma +  \frac{1}{d} \delta_{\mu\nu} h \, ,
\end{align}
where $\partial^2 = \partial^\lambda \partial_\lambda$ is the flat-space Euclidean Laplace operator. On the other hand, the analogous decomposition for torsion perturbations can be written as follows, where we are parameterizing the $1^+$ sector through transverse $2$-forms as in \cite{Martini:2023rnv}
\begin{align}\label{eq:new-york}
\delta T^\rho{}_{\mu\nu} = & \frac{1}{d-1} \left[ \delta^\rho{}_\nu \left( \tau_\mu + \partial_\mu \phi \right) - \delta^\rho{}_\mu \left( \tau_\nu + \partial_\nu \phi \right) \right] + H^\rho{}_{\mu\nu} + \partial^\rho \Pi_{\mu\nu} + \partial_\mu \Pi_\nu{}^\rho + \partial_\nu \Pi^\rho{}_\mu \\\nonumber
& + \kappa^\rho{}_{\mu\nu} + \partial_\mu \Sigma^\rho{}_\nu - \partial_\nu \Sigma^\rho{}_\mu + 2 \partial^\rho B_{\mu\nu} + \partial_\mu B^\rho{}_\nu - \partial_\nu B^\rho{}_\mu\\\nonumber
& + \partial^\rho \partial_\mu \zeta_\nu - \partial^\rho \partial_\nu \zeta_\mu + \frac{1}{d-1} \left( \delta^\rho{}_\nu \partial^2 \zeta_\mu - \delta^\rho{}_\mu \partial^2 \zeta_\nu \right) \, .
\end{align}
All the tensors that appear on the r.h.s.\ of the two previous equations are both transverse and traceless. Of course, one could also opt for a transverse decomposition as it was done in \cite{Baldazzi:2021kaf}. In this case, the infinitesimal gauge transformations of each spin-parity eigenstate would be different, but the physical properties of the gauge-invariant states would be the same. The spin-parity content of the previous decomposition is summarized in the Table \ref{table:1}.
\begin{table}[h!]
	\centering
	\begin{tabular}{||c | c | c | c || c | c | c | c ||} 
		\hline
		tensor & spin-parity & d.o.f. & irrep & tensor & spin-parity & d.o.f. & irrep \\ [0.5ex] \hline
		$\tau_\mu$ & $1^-$ & 3 & TrHA & $\kappa^\rho{}_{[\mu\nu]}$ & $2^-$ & 5 & TFHA \\ 
		\hline
		$\phi$ & $0^+$ & 1 & TrHA & $\Sigma_{(\mu\nu)}$ & $2^+$ & 5 & TFHA \\
		\hline
		$H_{[\rho\mu\nu]}$ & $0^-$ & 1 & TA & $B_{[\mu\nu]}$ & $1^+$ & 3 & TFHA \\
		\hline
		$\Pi_{[\mu\nu]}$ & $1^+$ & 3 & TA & $\zeta_\mu$ & $1^-$ & 3 & TFHA \\  
		\hline
	\end{tabular}
	\caption{Parameterization of the spin-parity content of torsion perturbations and their dimension in $d=4$. The last column identifies the $O(d)$-irreducible part of the torsion to which they concur. Indeed, ``HA'' stands for hook-antisymmetric, while ``TA'' means totally antisymmetric. Moreover, the prefix ``Tr'' means trace part, whereas ``TF'' indicates trace free.}
	\label{table:1}
\end{table}

The theory that we will consider in Sect.\ \ref{sect:inv-act} is built to be invariant under both diffeomorphisms and the local transformation \eqref{eq:deltaG}. In the flat-space limit, the representations of these gauge algebras acting on the metric and torsion are, respectively
\begin{subequations}
	\begin{align}
	& \delta^{\rm diff}_\epsilon h_{\mu\nu} = \partial_\mu \epsilon_\nu + \partial_\nu \epsilon_\mu \, ; \\\label{eq:flat-sp-gauge-A}
	& \delta^G_A T^\rho{}_{\mu\nu} = 3 \partial^\rho A_{\mu\nu} + \partial_\mu A^\rho{}_\nu - \partial_\nu A^\rho{}_\mu \, .
	\end{align}
\end{subequations}
In order to single out the gauge-invariant sector of the theory we further decompose the two tensors that parameterize these transformations into their transverse and longitudinal components, i.e.,
\begin{subequations}
	\begin{align}
	& \epsilon_\mu = \epsilon^T_\mu + \partial_\mu \varphi \, ;\\
	& A_{\mu\nu} = A^T_{\mu\nu} + \partial_\mu \xi^T_\nu - \partial_\nu \xi^T_\mu \, .
	\end{align}
\end{subequations}
As far as metric perturbations and infinitesimal diffeomorphisms are concerned we find the well known result that the $1^-$ d.o.f. is pure gauge, and that only one scalar mode is gauge invariant. The linear combination of the latter reads
\begin{equation}
s = h - \partial^2 \sigma \, .
\end{equation}
Now we focus on the second gauge transformation of our theory. To single out how the spin-parity eigenstates get shifted, we contract \eqref{eq:flat-sp-gauge-A} with the totally antisymmetric and hook antisymmetric algebraic projectors (see, e.g., \cite{Percacci:2020bzf}). In this way we read off the transformations in the $1^+$-sector
\begin{subequations}\label{eqs:gaugeA-1+}
	\begin{align}
	& B_{\mu\nu} \rightarrow B_{\mu\nu} + \frac{4}{3} A^T_{\mu\nu} \, ; \\
	& \Pi_{\mu\nu} \rightarrow \Pi_{\mu\nu} + \frac{1}{3} A^T_{\mu\nu} \, .
	\end{align}
\end{subequations}
Therefore, by performing the following rotation
\begin{subequations}\label{eqs:gaugeA-inv1+}
	\begin{align}
	& C_{\mu\nu} = \frac{4}{\sqrt{17}} \left( \frac{1}{4} B_{\mu\nu} - \Pi_{\mu\nu} \right) \, ;\\
	& F_{\mu\nu} = \frac{4}{\sqrt{17}} \left( \frac{1}{4} \Pi_{\mu\nu} + B_{\mu\nu} \right) \, ,
	\end{align}
\end{subequations}
we observe that $C_{\mu\nu}$ provides the gauge invariant physical mode.

We turn to the longitudinal part of the gauge transformation. Taking the trace of Eq.\ \eqref{eq:flat-sp-gauge-A} and projecting onto the tracefree hook antisymmetric part of the torsion we find
\begin{subequations}\label{eqs:gaugeA-1-}
	\begin{align}
	& \zeta_\mu \rightarrow \zeta_\mu + 4 \xi^T_\mu \, ; \\
	& \tau_\mu \rightarrow \tau_\mu - 4 \partial^2 \xi^T_\mu \, .
	\end{align}
\end{subequations}
Thus, the physical $1^-$ mode is
\begin{equation}\label{eq:gaugeA-inv-1-}
v_\mu = \zeta_\mu + \partial^2 \tau_\mu \, .
\end{equation}
As we were expecting from the onset, since the gauge invariance that we imposed is parameterized by an antisymmetric $2$-tensor, we have removed a transverse vector and a transverse $2$-form from the spectrum.

In the following we shall insert the spin-parity decompositions in the formal definitions of the quantum functional measures.

\subsection{Functional measure and non-local field redefinitions}\label{subsect:funct-measure-and-non-loc}

Here we introduce the functional measures of the metric and torsion fluctuations over which the path integral may be computed. Since these topics have already been dealt with in \cite{Mazur:1990ak} and \cite{Martini:2023rnv}, we will only review the main points that relevant for our discussion of the stability of the theory in Sect.\ \ref{sect:particle-content}.

The functional measure is usually defined through the normalization condition of the Gaussian ultra-local integral \cite{Mazur:1990ak,Mottola:1995sj,Percacci:2017fkn}. As we deal with the metric fluctuations we have \cite{Mazur:1990ak}
\begin{equation}\label{eq:def-funct-measure-metric}
\int {\cal D} h_{\mu\nu} {\rm e}^{- \int \sqrt{g} \, h \cdot G_h \cdot h} = 1 \, ,
\end{equation}
where $G_h$ is the Wheeler-DeWitt supermetric, whose explicit form can be parameterized as
\begin{equation}\label{eq:wheeler-supermetric}
 {G_h}^{\mu\nu\alpha\beta} = \frac{1}{2} \left[ g^{\mu\alpha} g^{\nu\beta} + g^{\mu\beta} g^{\nu\alpha} + c g^{\mu\nu} g^{\alpha\beta} \right] \, . 
\end{equation}
The constant $c$ is arbitrary, and it is a generic feature of integrating the fluctuations of tensors with rank $r \geq 2$. Indeed, the number of such constants is always given by $l-1$, where $l$ is the number of $SO(d)$-irreducible components of a given tensor. As a further step, one usually performs a change of variable in field space, in order to integrate over the spin-parity eigenstates that enter the covariant decomposition \eqref{eq:york}, i.e.,
\begin{equation}
{\cal D} h_{\mu\nu} = {\det({\cal J}_h)} {\cal D} h_{\mu\nu} \, {\cal D} \xi_{\mu} \, {\cal D} \sigma \, {\cal D} h .
\end{equation}
Here ${\cal J}_h$ is the Jacobian of the change of variable, whose determinant is found by plugging this last equation into Eq.\ \eqref{eq:def-funct-measure-metric} and employing Eq.\ \eqref{eq:wheeler-supermetric}
\begin{align}\label{eq:det-funct-jac-metric}
{\det({\cal J}_{h})} = \det \left( - 2 \partial^2 \right)^{1/2}\big|_{\xi_{\mu}} \, \det \left( 2 \frac{d-1}{d} \partial^4 \right)^{1/2}\bigg|_{\sigma} .
\end{align}

Now we turn to torsion fluctuations, and we start by defining the functional measure for in complete analogy with Eq.\ \eqref{eq:wheeler-supermetric} (see \cite{Martini:2023rnv})
\begin{equation}\label{eq:def-funct-measure-tor}
 \int {\cal D} T^\rho{}_{\mu\nu} {\rm e}^{- \int \sqrt{g} \, T \cdot G_T \cdot T} = 1 \, .
\end{equation}
In this case the exponential is the contraction of the torsion with the ultra-local supermetric for vector-valued $2$-forms. The explicit form of $G_T$ can be parameterized as
\begin{align}\label{eq:supermetric}\nonumber
 & {G_T}_\lambda{}^{\alpha\beta}{}_\rho{}^{\mu\nu} = \frac{1}{2} \left[ g_{\lambda\rho} \left( g^{\alpha\mu} g^{\beta\nu} - g^{\alpha\nu} g^{\beta\mu} \right) + \frac{a}{2} \left( \delta^\alpha{}_\lambda \delta^\mu{}_\rho g^{\beta\nu} - \delta^\beta{}_\lambda \delta^\mu{}_\rho g^{\alpha\nu} - \delta^\alpha{}_\lambda \delta^\nu{}_\rho g^{\beta\mu} + \delta^\beta{}_\lambda \delta^\nu{}_\rho g^{\alpha\mu} \right) \right. \\
 & \left. + \, 3 b  \left( g^{\alpha\mu} g^{\beta\nu} g_{\lambda\rho} - g^{\alpha\nu} g^{\beta\mu} g_{\lambda\rho} + \delta^\alpha{}_\rho \delta^\nu{}_\lambda g^{\beta\mu} - \delta^\alpha{}_\rho \delta^\mu{}_\lambda g^{\beta\nu} + \delta^\beta{}_\rho \delta^\mu{}_\lambda g^{\alpha\nu} - \delta^\beta{}_\rho \delta^\nu{}_\lambda g^{\alpha\mu} \right) \right] \, ,
\end{align}
where $a$ and $b$ are two arbitrary constants. Now we have one more constant w.r.t.\ Eq.\ \eqref{eq:wheeler-supermetric} we have one more algebraically irreducible component. Next, we split the functional measure over the spin-parity eigenstates as
\begin{equation}
 {\cal D} T^\rho{}_{\mu\nu} = {\det({\cal J}_T)} {\cal D} \kappa^\rho{}_{\mu\nu} \, {\cal D} H^\rho{}_{\mu\nu} \, {\cal D} S_{\mu\nu} \, {\cal D} A_{\mu\nu} \, {\cal D} \Pi_{\mu\nu} \, {\cal D} \tau_\mu \, {\cal D} \zeta_\mu \, {\cal D} \phi \, .
\end{equation}
In analogy with the previous case, ${\det(\cal J)}$ is the determinant of the Jacobian of the change of variable. Again, by plugging the decomposition \eqref{eq:new-york} into Eq.\ \eqref{eq:def-funct-measure-tor} we find
\begin{align}\label{eq:det-funct-jac}
 {\det({\cal J}_T)} = & \det \left( - 2 \partial^2 \right)^{1/2}\big|_{S_{\mu\nu}} \, \det \left( - 6 \partial^2 \right)^{1/2}\big|_{A_{\mu\nu}} \, \det \left( 2 \partial^4 \right)^{1/2}\big|_{\zeta_\mu} \,\\\nonumber
 & \det \left( - 3(1+b) \partial^2 \right)^{1/2}\big|_{\Pi_{\mu\nu}} \, \det \left( - \frac{2+3a}{3} \partial^2 \right)^{1/2}\bigg|_\phi \, .
\end{align}

Since the Laplacian $\partial^2$ is a negative semi-definite operator on Euclidean space, the functional determinants are well-defined only when powers of $-\partial^2$ are to be evaluated. Indeed, our condition for the absence of ghosts in Sect.\ \ref{sect:particle-content} exactly amounts to asking for the well-definiteness of these functional determinants.

From the last line in Eq.\ \eqref{eq:det-funct-jac} we obtain a constraint on the two arbitrary parameters that enter the supermetric \eqref{eq:supermetric}, that stem from requiring the positive definiteness of the operators, i.e.,
\begin{eqnarray}
 a>-\frac{2}{3} \, , && \qquad b>-1 \, .
\end{eqnarray}
Notice that we did not find any such constraint when analyzing from Eq.\ \eqref{eq:det-funct-jac-metric}. This is because the longitudinal modes parameterized by $\xi_\mu$ and $\sigma$ are orthogonal to the trace part of the Wheeler-DeWitt supermetric, which is proportional to $c$. Contrary, the scalar $\phi$ and the Kalb-Ramond field $\Pi_{\mu\nu}$ belong to the traceful and completely antisymmetric parts of the torsion, respectively, and the $a$ and $b$ constants tune the eigenvalues of these algebraically irreducible parts, see \cite{Martini:2023rnv}.

Let us consider a given a power-counting renormalizable Lagrangian density $\cal L$ for the torsion and metric, which involves kinetic terms of the torsion. Then, let $\cal L_{\rm flat}$ be the quadratic action that follows from expanding the fields around flat-space and retaining only quadratic terms. We can then decompose the torsion and metric according to their spin-parity decompositions, and functionally differentiate the result w.r.t.\ the spin-parity eigenfields. This procedure yields a block-diagonal Hessian, in which mixing is found only for fixed spin and parity. The Hessian is a differential operator whose order is greater or equal than two. In particular, for the spin-parity sectors pertaining to the torsion the order is equal to $2^{1+L}$, where $L$ is the number of longitudinal components. Then, we can combine the inverse of the square root of the functional determinant of the Hessian with Eq.\ \eqref{eq:det-funct-jac} to effectively lower the rank of the final differential operator. This procedure practically amounts to performing some non-local field redefinitions, and it is the strategy that we are going to follow in subsection \ref{subsect:particle-content} to work out the stability analysis of Sect.\ \ref{subsect:particle-content}. We remark that the necessity of performing non-local field redefinitions comes from having a non-trivial functional Jacobian, and that generic field redefinitions are generally not allowed in a classical field theory.

\section{Invariant action}\label{sect:inv-act}

In this section we shall derive the action that is invariant under the infinitesimal transformations \eqref{eq:deltaG}. In the construction of this action we will only keep power-counting renormalizable terms up to dimension four, i.e., up to $T^4$. The most general such action comprises $92$ independent scalar contractions \cite{Baldazzi:2021kaf}, involving also couplings with the Riemannian structure. In what follows we will not provide the expression of this general Lagrange density, and we refer to \cite{Baldazzi:2021kaf} for the counting and cataloguing of the different contractions\footnote{These terms can be found by applying the functions \texttt{AllContractions} and \texttt{MakeAnsatz} developed in xTras \cite{Nutma:2013zea}, and taking into account Bianchi identities and boundary terms.}. Moreover, due to the huge number of scalar contractions involved, all the calculations were done employing some of the packages that can be found in \texttt{xAct} \cite{Martin-Garcia:xAct}. The step-by-step derivation of the invariant action which follows the recipe that we outline below is relegated to the Appendix \ref{sect:app-derivation-action} to avoid overburdening the main body of the paper.

For the sake of the derivation it is convenient to split the transformation into its affine and homogeneous parts
\begin{equation}
 \delta^G = \delta^G_{AFF} + \delta^G_{HOM} \, .
\end{equation}
Concretely, we have that
\begin{subequations}
 \begin{align}
 & \delta^G_{AFF} T^\rho{}_{\mu\nu} = 3 \nabla^\rho A_{\mu\nu} + \nabla_\mu A^\rho{}_\nu - \nabla_\nu A^\rho{}_\mu \, ;\\
 & \delta^G_{HOM} T^\rho{}_{\mu\nu} = \zeta^2 \left[ 2 T_{[\mu\nu]}{}^\lambda A^\rho{}_\lambda + \frac{2}{3} T_{[\mu}{}^{\rho\lambda} A_{\nu]\lambda} + \frac{2}{3} T^\lambda{}_{\mu\nu} A^\rho{}_\lambda + \frac{4}{3} T^\lambda{}_{[\mu}{}^\rho A_{\nu]\lambda} - \frac{10}{3} T^\rho{}_{[\mu}{}^\lambda A_{\nu]\lambda} \right] \, .
 \end{align}
\end{subequations}
Then, let us split the most general action into the sum of terms $S_{k}$ that are homogeneous of degree $k$ in the torsion, i.e.,
\begin{equation}
 S[g,T] = \sum_{k=1}^{4} S_{k} [g,T] \, .
\end{equation}
Accordingly, to obtain a symmetric action we must have that
\begin{equation}\label{eq:aff-k-hom-k-1}
 \delta^G_{HOM} S_{k-1} + \delta^G_{AFF} S_k = 0 \, \quad \forall \quad  k \in \{1,2,3,4\} \, ,
\end{equation}
By hypothesis $k=4$ is the maximum degree is, thence we must have that
\begin{equation}
 \delta^G_{HOM} S_{4} = 0 \, .
\end{equation}
This equation yields conditions on the $33$ coefficients that enter $S_4$. On the other hand, by applying Eq. \eqref{eq:aff-k-hom-k-1} with $k=4$ we obtain conditions on the $31$ couplings entering $S_3$ in terms of those present in $S_4$. This iterative algorithm stops at $k=1$, when we must have
\begin{equation}
 \delta^G_{AFF} S_1 = 0 \, .
\end{equation}
Clearly, the validity of this last equation is a consistency check that may single out non-invariant actions. Nevertheless, due to the tensor structure of the transformation \eqref{eq:deltaG} and the second Bianchi identities this last equation holds identically. Thus, it does not further constrain the coefficients of the starting action.

In finding the solutions we have discarded those that fix to zero the parameter $\zeta$ which tunes $\delta^G_{HOM}$. Besides the trivial solution we find a non-trivial one, which can be written in terms of a single coupling constant $\xi$ and the parameter $\zeta$. Thus, by imposing the invariance of the full action under \eqref{eq:deltaG}, we have fixed the values of $90$ of the $92$ coupling constants.

The importance of such a dramatic decrease in the parameter space of the theory cannot be overstated. Indeed, this paves the way for working out sensible predictions out of such a torsionful theory of gravity. Moreover, only the ratio of the two constants $\xi$ and $\zeta$ appears in the action. To present this symmetric solution we write the action as the sum of the homogeneous parts once again, i.e.,
\begin{equation}\label{eq:inv-act}
 S_{\rm inv}[g,T] = S_{1}[g,T] + S_{2}[g,T] + S_{3}[g,T] + S_{4}[g,T] \, .
\end{equation}
The first term on the right-hand-side contains only one of the two possible terms, and it reads
\begin{equation}\label{eq:inv-act-o1}
S_1[g,T] = - \frac{2\xi}{\zeta} \int \sqrt{g} T^{\rho\mu\nu} \nabla_\mu R_{\nu\rho} \, .
\end{equation}
When it comes to quadratic terms we notice that none of the mass terms for the torsion is compatible with $\delta^0 S=0$, i.e., the torsion sector scale-invariant. Therefore, the full expression of the quadratic part of the action is
\begin{align}\label{eq:inv-act-o2}
 S_2[g,T] = & - \int \sqrt{g} \left[ - (\nabla_\alpha T_{\mu\nu\rho}) \nabla^\alpha T^{\mu\nu\rho} + \frac{5}{3} (\nabla_\alpha T_{\mu\nu\rho}) \nabla^\alpha T^{\nu\mu\rho} + \frac{2}{3} (\nabla_\lambda T^{\lambda\mu\nu}) \nabla_\alpha T^\alpha{}_{\mu\nu} \right.\\\nonumber
 & \left. - \frac{1}{3} (\nabla_\alpha T^{\mu\alpha\nu}) \nabla_\lambda T_\nu{}^\lambda{}_\mu + \frac{4}{3} (\nabla_\alpha T^{\alpha\mu\nu}) \nabla_\alpha T_{\mu\nu}{}^\alpha + \frac{1}{3} (\nabla_\lambda T^{\mu\lambda\nu}) \nabla_\alpha T_\mu{}^\alpha{}_\nu + \frac{2}{3} T^{\mu\alpha\nu} T^\rho{}_\alpha{}^\sigma R_{\mu\rho\nu\sigma} \right.\\\nonumber
 & \left. + \frac{8}{3} T^{\mu\alpha\nu} T^\rho{}_\alpha{}^\sigma R_{\mu\nu\rho\sigma} + \frac{1}{3} T^{\lambda\mu\nu} T_\lambda^{\alpha\beta} R_{\mu\nu\alpha\beta} - 2 T^{\alpha\lambda\beta} T_\lambda{}^{\mu\nu} R_{\alpha\beta\mu\nu} + \frac{4}{3} T^{\alpha\mu\nu} T_\mu{}^\beta{}_\nu R_{\alpha\beta} \right.\\\nonumber
 & \left. - \frac{2}{3} T^{\alpha\mu\nu} T^\beta{}_{\mu\nu} R_{\alpha\beta} + \frac{2}{3} T^{\rho\alpha\sigma} T_{[\sigma}{}^\beta{}_{\rho]} R_{\alpha\beta} \right] \, .
\end{align}
Moreover, we notice that the previous expression displays only $6$ out of the $9$ kinetic terms of the torsion, and $8$ out of the $14$ contractions with the Riemann and Ricci tensor and the Ricci scalar. Turning to cubic terms, we obtain
\begin{align}\label{eq:inv-act-o3}
& S_3[g,T] = - \frac{\zeta}{\xi} \int \sqrt{g} \left[ \frac{4}{9} T^{\rho\mu\nu} T_\mu{}^{\alpha\beta} \nabla_\rho T_{\nu\alpha\beta} - \frac{8}{3} T^{\rho\mu\nu} T^\alpha{}_\mu{}^\beta \nabla_\rho T_{\nu\alpha\beta} - \frac{4}{3} T^{\rho\mu\nu} T_\mu{}^{\alpha\beta} \nabla_\nu T_{\rho\alpha\beta} \right.\\\nonumber
& \left. + \frac{4}{3} T^{\rho\mu\nu} T^\alpha{}_\mu{}^\beta \nabla_\nu T_{\rho\alpha\beta} - \frac{20}{9} T^{\rho\mu\nu} T^\alpha{}_\mu{}^\beta \nabla_\alpha T_{\rho\nu\beta} + \frac{16}{9} T^{\rho\mu\nu} T^\alpha{}_\mu{}^\beta \nabla_\alpha T_{\nu\rho\beta} - \frac{20}{9} T^{\rho\mu\nu} T_\mu{}^{\alpha\beta} \nabla_\beta T_{\rho\nu\alpha} \right.\\\nonumber
& \left. - \frac{2}{3} T^{\rho\mu\nu} T_{\mu\nu}^\alpha \nabla_\beta T_{\rho\alpha}{}^\beta - \frac{2}{9} T^{\rho\mu\nu} T^\alpha{}_{\mu\nu} \nabla_\beta T_{\rho\alpha}{}^\beta - \frac{16}{9} T^{\rho\mu\nu} T^\alpha{}_\mu{}^\beta \nabla_\beta T_{\nu\rho\alpha} + \frac{10}{9} T^{\rho\mu\nu} T_{\rho\mu}{}^\alpha \nabla_\beta T_{\alpha\nu}{}^\beta \right.\\\nonumber
& \left. + \frac{8}{9} T^{\rho\mu\nu} T_{\mu\rho}{}^\alpha \nabla_\beta T_{\nu\alpha}{}^\beta - \frac{4}{3} T^{\rho\mu\nu} T_\mu{}^{\alpha\beta} \nabla_\beta T_{\alpha\rho\nu} + \frac{2}{3} T^{\rho\mu\nu} T_{\mu\nu}{}^\alpha \nabla_\beta T_{\alpha\rho}{}^\beta - \frac{4}{3} T^{\rho\mu\nu}  T_{\mu\nu}{}^\alpha \nabla_\beta T^\beta{}_{\rho\alpha} \right] \, .
\end{align}
In this case we observe that only $14$ out of the $31$ independent contractions actually enter the Lagrangian. Lastly, the quartic part of the invariant action comprises $17$ of the $33$ independent scalars, i.e.,
\begin{align}\label{eq:inv-act-o4}\nonumber
& S_4[g,T] = - \frac{\zeta^2}{\xi^2} \int \sqrt{g} \left[ \frac{16}{27} T_\rho{}^{\mu\nu} T^{\rho\alpha\beta} T_{\alpha\mu}{}^\lambda T_{\beta\nu\lambda} - \frac{1}{3} T_\rho{}^{\mu\nu} T^{\rho\alpha\beta} T_{\alpha\beta}{}^\lambda T_{\mu\nu\lambda} - \frac{5}{27} T_\rho{}^{\mu\nu} T^\rho{}_\mu{}^\alpha T_\alpha{}^{\beta\lambda} T_{\nu\beta\lambda} \right.\\\nonumber
& \left. - \frac{32}{27} T^{\rho\mu\nu} T_\mu{}^{\alpha\beta} T_{\nu\alpha}{}^\lambda T_{\beta\rho\lambda} + \frac{1}{3} T^{\rho\mu\nu} T_{\mu\nu}{}^\alpha T_\alpha{}^{\beta\lambda} T_{\beta\rho\lambda} - \frac{2}{3} T_\rho{}^{\mu\nu} T^{\rho\alpha\beta} T_{\alpha\mu}{}^\lambda T_{\nu\beta\lambda} + \frac{10}{9} T_{\rho\mu}{}^\nu T^{\rho\mu\alpha} T_\alpha{}^{\beta\lambda} T_{\beta\nu\lambda} \right.\\\nonumber
& \left. - \frac{8}{9} T^{\rho\mu\nu} T_{\mu\rho}{}^\alpha T_\nu{}^{\beta\lambda} T_{\beta\alpha\lambda} - \frac{25}{54} T^{\rho\mu\nu} T_{\rho\mu}{}^\alpha T^\beta{}_\nu{}^\lambda T_{\beta\alpha\lambda} - \frac{1}{3} T^{\rho\mu\nu} T_\mu{}^{\alpha\beta} T_{\alpha\nu}{}^\lambda T_{\lambda\rho\beta} + \frac{44}{27} T^{\rho\mu\nu} T_\rho{}^{\alpha\beta} T_{\mu\alpha}{}^\lambda T_{\lambda\nu\beta} \right.\\\nonumber
& \left. - \frac{2}{9} T_\rho{}^{\mu\nu} T^{\rho\alpha\beta} T_{\alpha\beta}{}^\lambda T_{\lambda\mu\nu} + \frac{20}{27} T_{\rho\mu}{}^\alpha T^{\rho\mu\nu} T^\beta{}_\nu{}^\lambda T_{\lambda\alpha\beta} - \frac{8}{27} T^{\rho\mu\nu} T_{\mu\rho}{}^\alpha T^\beta{}_\nu{}^\lambda T_{\lambda\alpha\beta} - \frac{1}{54} T_\rho{}^{\mu\nu} T^{\rho\alpha\beta} T_{\lambda\mu\nu} T^\lambda{}_{\alpha\beta} \right.\\
& \left. + \frac{1}{3} T_\rho{}^{\mu\nu} T^{\rho\alpha\beta} T_{\lambda\beta\nu} T^\lambda{}_{\alpha\mu} + \frac{4}{27} T^{\rho\mu\nu} T_{\mu\rho}{}^\alpha T_\nu{}^{\beta\lambda} T_{\alpha\beta\lambda} \right] \, .
\end{align}
When $\frac{\xi}{\zeta} \ll 1$ the term linear in the torsion in Eq.\ \eqref{eq:inv-act-o1} can be treated perturbatively, whereas we cannot do the same with the cubic and quartic ones.

In order to better grasp the structure of these interaction terms, we can proceed to perform a partial algebraic decomposition of the torsion singling out the completely antisymmetric part, i.e.,
\begin{equation}
	T^\rho{}_{\mu\nu} = t^\rho{}_{\mu\nu} + \varepsilon^\rho{}_{\mu\nu\lambda} \theta^\lambda \, ,
\end{equation}
where $t^\rho{}_{\mu\nu}$ is hook antisymmetric and has a nontrivial trace. The first part of the interaction that we consider is that which is quartic in $t^\rho{}_{\mu\nu}$, and it reads
\begin{equation}
	\begin{split}
		{\cal L}_{\rm quar} = \frac{\zeta^2}{\xi^2} & \left[ - \tfrac{16}{27} t_{\alpha }{}^{\rho \lambda} t^{\alpha \mu \nu} t_{\mu \rho}{}^{\sigma} t_{\nu \lambda \sigma} -  \tfrac{4}{27} t_{\alpha \mu}{}^{\rho} t^{\alpha \mu \nu} t_{\nu}{}^{\lambda \sigma} t_{\rho \lambda \sigma} + \tfrac{32}{27} t^{\alpha \mu \nu} t_{\mu}{}^{\rho \lambda} t_{\nu \rho}{}^{\sigma} t_{\lambda \alpha \sigma} \right.\\
		& \left. + \tfrac{2}{3} t_{\alpha }{}^{\rho \lambda} t^{\alpha \mu \nu} t_{\mu \rho}{}^{\sigma} t_{\lambda \nu \sigma} + \tfrac{1}{54} t_{\alpha \mu}{}^{\rho} t^{\alpha \mu \nu} t_{\lambda \rho \sigma} t^{\lambda}{}_{\nu}{}^{\sigma} + \tfrac{1}{3} t^{\alpha \mu \nu} t_{\mu}{}^{\rho \lambda} t_{\rho \nu}{}^{\sigma} t_{\sigma \alpha \lambda} \right.\\
		& \left. -  \tfrac{44}{27} t_{\alpha }{}^{\rho \lambda} t^{\alpha \mu \nu} t_{\mu \rho}{}^{\sigma} t_{\sigma \nu \lambda} + \tfrac{8}{27} t_{\alpha }{}^{\rho \lambda} t^{\alpha \mu \nu} t_{\sigma \rho \lambda} t^{\sigma}{}_{\mu \nu} -  \tfrac{1}{3} t_{\alpha }{}^{\rho \lambda} t^{\alpha \mu \nu} t_{\sigma \nu \lambda} t^{\sigma}{}_{\mu \rho} \right] \, .
	\end{split}
\end{equation}
On the other hand, the terms linear in the axial torsion $\theta^\mu$
\begin{equation}
	\begin{split}
		{\cal L}_{\rm quar} = \frac{\zeta^2}{\xi^2} & \left[ \tfrac{112}{27} \varepsilon_{\alpha \rho \sigma \beta} t_{\mu}{}^{\lambda \sigma} t^{\mu \nu \rho} t_{\nu \lambda}{}^{\beta} \theta^{\alpha } + \tfrac{16}{9} \varepsilon_{\alpha \mu \sigma \beta} t^{\mu \nu \rho} t_{\nu}{}^{\lambda \sigma} t_{\rho \lambda}{}^{\beta} \theta^{\alpha } + \tfrac{4}{27} \varepsilon_{\alpha \lambda \sigma \beta} t_{\mu \nu}{}^{\lambda} t^{\mu \nu \rho} t_{\rho}{}^{\sigma \beta} \theta^{\alpha } \right.\\
		& \left. + \tfrac{32}{27} \varepsilon_{\alpha \rho \sigma \beta} t^{\mu \nu \rho} t_{\nu}{}^{\lambda \sigma} t_{\lambda \mu}{}^{\beta} \theta^{\alpha } -  \tfrac{112}{27} \varepsilon_{\alpha \mu \sigma \beta} t^{\mu \nu \rho} t_{\nu}{}^{\lambda \sigma} t_{\lambda \rho}{}^{\beta} \theta^{\alpha } -  \tfrac{16}{27} \varepsilon_{\alpha \lambda \sigma \beta} t_{\mu}{}^{\lambda \sigma} t^{\mu \nu \rho} t^{\beta}{}_{\nu \rho} \theta^{\alpha } \right.\\
		& \left. + \tfrac{80}{27} \varepsilon_{\alpha \rho \sigma \beta} t_{\mu}{}^{\lambda \sigma} t^{\mu \nu \rho} t^{\beta}{}_{\nu \lambda} \theta^{\alpha } + \tfrac{112}{27} \varepsilon_{\alpha \mu \sigma \beta} t^{\mu \nu \rho} t_{\nu}{}^{\lambda \sigma} t^{\beta}{}_{\rho \lambda} \theta^{\alpha } + \tfrac{16}{27} \varepsilon_{\alpha \mu \lambda \beta} t^{\mu \nu \rho} t^{\lambda}{}_{\nu}{}^{\sigma} t^{\beta}{}_{\rho \sigma} \theta^{\alpha } \right] \, .
	\end{split}
\end{equation}
No term which is cubic in the axial can be written down, therefore we are only left with those that are quadratic and quartic in $\theta^\mu$, which are given by
\begin{equation}
	\begin{split}
		{\cal L}_{\rm quar} = \frac{\zeta^2}{\xi^2} & \left[ -  \tfrac{128}{27} t_{\mu \nu \rho} t^{\mu \nu \rho} \theta_{\alpha } \theta^{\alpha } -  \tfrac{16}{27} t^{\mu}{}_{\mu}{}^{\nu} t^{\rho}{}_{\nu \rho} \theta_{\alpha } \theta^{\alpha } + \tfrac{8}{3} t_{\alpha }{}^{\nu \rho} t_{\mu \nu \rho} \theta^{\alpha } \theta^{\mu} \right.\\
		& \left. + \tfrac{16}{27} t^{\nu}{}_{\alpha \nu} t^{\rho}{}_{\mu \rho} \theta^{\alpha } \theta^{\mu} + \tfrac{32}{27} t_{\alpha \mu}{}^{\nu} t^{\rho}{}_{\nu \rho} \theta^{\alpha } \theta^{\mu} + \tfrac{256}{9} \theta_{\alpha } \theta^{\alpha } \theta_{\mu} \theta^{\mu} \right] \, .
	\end{split}
\end{equation}

We have checked the conformal properties of the action, and it turns out that it is not conformal. This statement is valid for both weak and strong conformal symmetry (see \cite{Shapiro:2001rz}), the first case being for non-transforming torsion, while in the latter the torsion vector plays the role of a Weyl potential (see also \cite{Sauro:2022hoh}). To see that the theory is not conformal in the weak sense one can either perform the explicit calculation, or just observe that the infinitesimal Weyl transformation of \eqref{eq:inv-act-o1} cannot be counterbalanced by any other term in the action (see \cite{Erdmenger:1997wy} for the details of this reasoning). On the other hand, if the action were conformal in the strong sense, then there would be terms depending only on the torsion vector in \eqref{eq:inv-act-o3}.

To get a better insight into the GR limit of the previous action, let us compute the field equations of the torsion by performing a functional variation of the gauge invariant action Eq.\ \eqref{eq:inv-act}. To this end, we shall discard the precise tensor structure that one gets by varying Eq.s \eqref{eq:inv-act-o3} and \eqref{eq:inv-act-o4}. Instead, these are implicitly written as $f(T^2)_\rho{}^{\mu\nu}$ and $g(T^3)_\rho{}^{\mu\nu}$, respectively. Thus, by retaining explicitly only terms up to order one in the torsion the field equations are
\begin{align}
 & \frac{5}{3} \nabla_{\alpha }\nabla^{\alpha }T^{\nu \mu }{}_{\rho } -  \frac{5}{3} \nabla_{\alpha }\nabla^{\alpha }T^{\mu \nu }{}_{\rho } - 2 \nabla_{\alpha }\nabla^{\alpha }T_{\rho }{}^{\mu \nu } + \frac{2}{3} \nabla^{\mu }\nabla_{\alpha }T^{\alpha \nu }{}_{\rho } + \frac{1}{3} \nabla^{\mu 
 }\nabla_{\alpha }T^{\nu }{}_{\rho }{}^{\alpha } \\\nonumber
 &-  \frac{1}{3}  \nabla^{\mu }\nabla_{\alpha }T_{\rho }{}^{\nu \alpha } -\frac{2}{3} \nabla^{\nu }\nabla_{\alpha }T^{\alpha \mu  }{}_{\rho } -  \frac{1}{3} \nabla^{\nu }\nabla_{\alpha }T^{\mu }{}_{\rho }{}^{\alpha } + \frac{1}{3} \nabla^{\nu }\nabla_{\alpha }T_{\rho }{}^{\mu \alpha } + \frac{4}{3} \nabla_{\rho }\nabla_{\alpha }T^{\alpha \mu \nu } + \frac{2}{3} \nabla_{\rho }\nabla_{\alpha }T^{\mu \nu \alpha }\\\nonumber
 & -  \frac{2}{3} \nabla_{\rho }\nabla_{\alpha }T^{\nu \mu \alpha } +
 \frac{4}{3} R_{\rho }{}^{\alpha } T_{\alpha }{}^{\mu \nu } -  \frac{2}{3} R^{\nu \alpha } T_{\alpha }{}^{\mu }{}_{\rho } + \frac{2}{3} R^{\mu \alpha } T_{\alpha }{}^{\nu }{}_{\rho }\\\nonumber
 & -  \frac{2}{3} R^{\nu }{}_{\alpha_1 \rho \alpha } T^{\alpha \mu \alpha_1} +  \frac{8}{3} R^{\nu }{}_{\rho \alpha \alpha_1} T^{\alpha \mu \alpha_1} + \frac{2}{3} R^{\mu 
 }{}_{\alpha_1 \rho \alpha } T^{\alpha \nu \alpha_1} -  \frac{8}{3} R^{\mu }{}_{\rho \alpha \alpha_1}  T^{\alpha \nu \alpha_1} + 2 R^{\mu \nu }{}_{\alpha 
 	\alpha_1} T^{\alpha }{}_{\rho }{}^{\alpha_1}\\\nonumber
 & -  R^{\nu  }{}_{\rho \alpha \alpha_1} T^{\mu \alpha \alpha_1} +  \frac{2}{3} R_{\rho }{}^{\alpha } T^{\mu \nu  }{}_{\alpha } -  \frac{1}{3} R^{\nu \alpha } T^{\mu 
 }{}_{\rho \alpha } + R^{\mu }{}_{\rho \alpha \alpha_1}  T^{\nu \alpha \alpha_1} -  \frac{2}{3} R_{\rho  }{}^{\alpha } T^{\nu \mu }{}_{\alpha } + \frac{1}{3} 
 R^{\mu \alpha } T^{\nu }{}_{\rho \alpha } \\\nonumber
 & -  \frac{2}{3} R^{\mu \nu }{}_{\alpha \alpha_1} T_{\rho }{}^{\alpha \alpha_1} + \frac{1}{3} R^{\nu \alpha } T_{\rho }{}^{\mu }{}_{\alpha } -  \frac{1}{3} R^{\mu  	\alpha } T_{\rho }{}^{\nu }{}_{\alpha } + \frac{\zeta}{\xi} f(T^2)_\rho{}^{\mu\nu} + \frac{\zeta^2}{\xi^2} g(T^3)_\rho{}^{\mu\nu} 
 = \frac{\xi}{\zeta} \left(\nabla^{\mu }R^{\nu }{}_{\rho } - \nabla^{\nu }R^{\mu }{}_{\rho } \right) \, .
\end{align}
Therefore, in the limit of vanishing torsion we have that the antisymmetrized covariant derivative of the Ricci tensor is zero. This is clearly the case for all the solutions of the Einstein field equations, even if we introduce a non-trivial cosmological constant. Thus, the theory is classically compatible with General Relativity if we switch off the torsion. Moreover, we notice that a modification of the Cotton tensor appears as a source on the right-hand side of the field equations.

The invariant action \eqref{eq:inv-act} must be supplemented with the purely metric sector, which naturally comprises higher-derivative terms \cite{Stelle:1976gc,Stelle:1977ry}. This is because such terms would in any case be produced by radiative corrections, and therefore we must take them into account from the onset. Notice that the action Eq.\ \eqref{eq:inv-act} is invariant under the discrete transformation $g_{\mu\nu} \rightarrow - g_{\mu\nu}$. Requiring this discrete symmetry also in the purely metric sector would forbid the appearance of the Einstein-Hilbert term. Thus, since we are interested in recovering General Relativity in the low energy limit, we do not impose it. Accordingly, the full action functional whose stability we shall analyze in the Sect.\ \ref{subsect:particle-content} is
\begin{equation}\label{eq:full-action}
 S_{\rm full}[g,T] = S_{\rm inv}[g,T] - \int \sqrt{g} \left[ \alpha R_{\mu\nu}R^{\mu\nu} - \beta R^2 + \frac{2}{\varkappa^2} R \right] \, ,
\end{equation}
where we have chosen to adhere to the same convention used in \cite{Stelle:1977ry} for labeling the coupling constants and $\varkappa^2 = 16 \pi G$.

After having derived the action, in the following section we shall proceed to study the particle content of the theory.

\section{Spin-parity decomposed action and stability}\label{sect:particle-content}

In this section we apply the standard formalism of the spin-parity decomposition to the action \eqref{eq:full-action} in order to study the stability of the theory. To do so, we expand the action around flat-space retaining only quadratic terms, and then we apply the decompositions \eqref{eq:york} and \eqref{eq:new-york}. The terms that concur to the flat-space action are
\begin{align}\label{eq:full-action-quad}
 S_{\rm quad}[g,T] = - \int \sqrt{g} & \left[ - (\nabla_\alpha T_{\mu\nu\rho}) \nabla^\alpha T^{\mu\nu\rho} + \frac{5}{3} (\nabla_\alpha T_{\mu\nu\rho}) \nabla^\alpha T^{\nu\mu\rho} + \frac{2}{3} (\nabla_\lambda T^{\lambda\mu\nu}) \nabla_\alpha T^\alpha{}_{\mu\nu} \right.\\\nonumber
 & \left. \,\, - \frac{1}{3} (\nabla_\alpha T^{\mu\alpha\nu}) \nabla_\lambda T_\nu{}^\lambda{}_\mu + \frac{4}{3} (\nabla_\alpha T^{\alpha\mu\nu}) \nabla_\alpha T_{\mu\nu}{}^\alpha + \frac{1}{3} (\nabla_\lambda T^{\mu\lambda\nu}) \nabla_\alpha T_\mu{}^\alpha{}_\nu \right.\\\nonumber
 & \left. \,\, + \frac{2\xi}{\zeta} T^{\rho\mu\nu} \nabla_\mu R_{\nu\rho} + \alpha R_{\mu\nu}R^{\mu\nu} - \beta R^2 + \frac{2}{\varkappa^2} R \right] \, .
\end{align}
and comprise four undetermined couplings, beside $\varkappa^2 = 16 \pi G$. 

As we have anticipated in Sect.\ \ref{subsect:funct-measure-and-non-loc}, the kinetic terms of non-transverse modes are higher-derivative ones, and in Subsect.\ \ref{subsect:particle-content} we shall follow the method of \cite{Mazur:1990ak} to reduce their order.

\subsection{Spin-parity decomposed actions in the flat-space limit}\label{subsect:spin-parity-lagr}

Since in flat-space devoid of non-trivial background fields (beside the metric) different spin-parity sectors are orthogonal to each other, action is written as a sum over all the spins and parities, i.e.,
\begin{equation}\label{eq:action-quad-flat}
 S_{\small\rm 2flat} = \sum_{J,P} \int d^4x  \mathcal{L}_{\small J^{P}} \, .
\end{equation}
We will first analyze the $J^P$-sectors that pertain to the torsion only, focusing on others afterwards.

We commence our analysis by considering the two fully gauge-invariant sectors that are carried by the torsion, i.e., the $2^-$ and $0^-$. Their Lagrangians are of second-order, and they do not give rise to any Jacobian factor in the functional measure \eqref{eq:det-funct-jac}. Thus, we simply have
\begin{equation}\label{eq:flat-sp-lagr-2^-0^-}
  \mathcal{L}_{\small\rm 2^{-}} + \mathcal{L}_{\small\rm 0^{-}} =  - \frac{1}{6} \kappa^{\rho\mu\nu} \partial^2 \kappa_{\rho\mu\nu} - \frac{8}{3} H^{\rho\mu\nu} \partial^2 H_{\rho\mu\nu} \, .
\end{equation}
Since the are no contributions from these sectors in \eqref{eq:det-funct-jac}, we can already state that the theory is devoid of kinematical ghosts in these sectors.

Let us now analyze the $1^+$ sector, whose Lagrangian is necessarily of fourth-order due to the structure of the decomposition \eqref{eq:new-york}. Moreover, it is manifestly gauge-invariant under \eqref{eq:flat-sp-gauge-A}, resulting in
\begin{align}\label{eq:flat-sp-lagr-1+}
 \mathcal{L}_{\small\rm 1^{+}} = & \frac{1}{3} B^{\mu\nu} (\partial^2)^2 B_{\mu\nu} - \frac{8}{3} B^{\mu\nu} (\partial^2)^2 \Pi_{\mu\nu} + \frac{16}{3} \Pi^{\mu\nu} (\partial^2)^2 \Pi_{\mu\nu} \\\nonumber
 = & \frac{16}{3} C^{\mu\nu} (\partial^2)^2 C_{\mu\nu} \, ,
\end{align}
where in the second line we have inserted the gauge-invariant combination $C_{\mu\nu}$, see Eq.\ \eqref{eqs:gaugeA-inv1+}.

We turn to the $1^-$ sector, whose only gauge-invariant field variable pertains to the torsion. Indeed, the pure gauge transverse vector $\xi^\mu$ that appears in the York decomposition \eqref{eq:york} exactly drops out from the final expression. Furthermore, the flat-space Lagrangian is manifestly invariant under the linearized action of \eqref{eq:deltaG}, and it can be expressed in terms of the gauge-invariant vector $v^\mu$ only, see \eqref{eq:gaugeA-inv-1-}, i.e.,
\begin{align}\label{eq:flat-sp-lagr-1-}
 \mathcal{L}_{\small\rm 1^{-}} = & - \frac{2(d-2)}{3(d-1)^2} \left(\partial^2 \zeta^\mu + \tau^\mu \right) \partial^2 \left(\partial^2 \zeta_\mu + \tau_\mu \right) \\\nonumber
 = & - \frac{4}{27} v_\mu \partial^2 v^\mu \, ,
\end{align}
In the last line we have specialized to $d=4$.

Now, we consider the $2^+$ and $0^+$ sectors, in which we have mixing between the torsion and metric fluctuations, starting from the first one. Here the two field variables are the symmetric transverse traceless tensors $h_{\mu\nu}$ and $S_{\mu\nu}$, and the Lagrangian is
\begin{equation}\label{eq:flat-sp-lagr-2+}\nonumber
\mathcal{L}_{\small\rm 2^{+}} = \frac{2}{3} S^{\mu\nu} \partial^4 S_{\mu\nu} - \frac{1}{2} \alpha h_{\mu\nu} \partial^4 h^{\mu\nu}  - \frac{1}{\varkappa^2} h_{\mu\nu} \partial^2 h^{\mu\nu} - \frac{2\xi}{\zeta} h_{\mu\nu} \partial^4 S^{\mu\nu}  \, .
\end{equation}
In this case the Lagrangian displays both fourth- and second-order operators, and three out of the four free parameters of the theory enter it.

Eventually, let us consider the $0^+$ sector, whose degrees of freedom are described by the scalars $h$ and $\sigma$ that appear in the York decomposition of the metric \eqref{eq:york}, and by the scalar $\phi$ that enters Eq.\ \eqref{eq:new-york}. Again, the Lagrangian is gauge-invariant, allowing us to express the result in terms of the gauge invariant field $s$ only, i.e.,
\begin{align}\label{eq:flat-sp-lagr-0^+}
\mathcal{L}_{\small\rm 0^{+}} = & \frac{(d-1)(d-2)}{d^2 \, \varkappa^2} \, s  \partial^2 s + \frac{d-1}{2d^2} (4 (d-1) \beta - d \alpha ) \, s  \partial^4 s + \frac{2}{3(d-1)} \phi \partial^4 \phi + \frac{2 \zeta}{d \, \xi} \phi \partial^4 s \\\nonumber
= & \frac{3}{8 \varkappa^2} \, s  \partial^2 s + \frac{3}{8} (3 \beta - \alpha ) \, s  \partial^4 s + \frac{2}{9} \phi \partial^4 \phi + \frac{\xi}{2 \, \zeta} \phi \partial^4 s \, .
\end{align}
As in $2^+$ case, the differential operators also comprise a second-order contribution coming from the Einstein-Hilbert action. Moreover, this is the only spin-parity sector in which all the undetermined couplings of the theory appear.


\subsection{Stability analysis}\label{subsect:particle-content}

In this subsection we combine the results of the flat-space Lagrangians that we have just presented with the functional measure over torsion and metric fluctuations outlined in Subsect.\ \ref{subsect:funct-measure-and-non-loc}. This yields a partition function ${\cal Z}$ which is given by the product of the functional determinants over the transverse traceless tensors that parameterize all the spin-parity sectors. Concretely, ${\cal Z}$ is defined as
\begin{equation}
 {\cal Z} = \int {\cal D} T^\rho{}_{\mu\nu} \, {\cal D} h_{\mu\nu} \, {\rm e}^{- S_{\text{\tiny 2flat}}} = {\cal Z_{\text{\tiny $2^-$}}} {\cal Z_{\text{\tiny $2^+$}}} {\cal Z_{\text{\tiny $1^-$}}} {\cal Z_{\text{\tiny $1^+$}}} {\cal Z_{\text{\tiny $0^-$}}} {\cal Z_{\text{\tiny $0^+$}}} \, ,
\end{equation}
where $S_{\text{\tiny 2flat}}$ is given by \eqref{eq:action-quad-flat}.

We start from the $2^-$ and $0^-$ sectors, see Eq.\ \eqref{eq:flat-sp-lagr-2^-0^-}. In this case we do not have any non-trivial Jacobian to play with; thence, the contribution to the partition function of these modes is simply given by
\begin{equation}\label{eq:part-funct-2-0-}
 {\cal Z_{\text{\tiny $2^-$}}} {\cal Z_{\text{\tiny $0^-$}}} =  \int {\cal D} \kappa^\rho{}_{\mu\nu} {\cal D} H^\rho{}_{\mu\nu} {\rm e}^{- \int d^4x \left( {\cal L}_{\text{\tiny $2^{-}$}} + {\cal L}_{\text{\tiny $0^{-}$}}\right) } = \det \left( - \partial^2 \right)^{-1/2}\bigg|_{\kappa^\rho{}_{\mu\nu}} \, \det \left( - \partial^2 \right)^{-1/2}\bigg|_{H^\rho{}_{\mu\nu}} \, ,
\end{equation}
where we have discarded positive numerical factors multiplying the Laplacian by assuming the absence of multiplicative anomalies.
Since $-\partial^2$ is a positive semi-definite operator in Euclidian space, the functional determinant that are to be evaluated are negative semi-definite. Therefore, these spin-parity sectors yield meaningful results, and they do not give rise to ghost instabilities.

Now we turn to the $1^+$ sector. In this case, the Hessian is degenerate, c.f.\ \eqref{eq:flat-sp-lagr-1+}. By performing the rotation \eqref{eqs:gaugeA-inv1+} we re-write the Lagrangian in terms of the gauge-invariant field variable $C_{\mu\nu}$. Thus, the functional integral over $F_{\mu\nu}$ only involves the measure, and assuming $b>1$ and the absence of multiplicative anomalies we obtain
\begin{align}\label{eq:part-funct-1+}\nonumber
 {\cal Z_{\text{\tiny $1^+$}}} \supset & \int {\cal D} B_{\mu\nu} {\cal D} \Pi_{\mu\nu} \det \left( - 6 \partial^2 \right)^{1/2}\big|_{B_{\mu\nu}} \det \left( - 3(1+b) \partial^2 \right)^{1/2}\big|_{\Pi_{\mu\nu}}\, {\rm e}^{- \int d^4x \left( {\cal L}_{\text{\tiny $1^{+}$}} \right)}\\\nonumber
 = & \int {\cal D} C_{\mu\nu} {\cal D} F_{\mu\nu} \det \left( - \partial^2 \right)^{1/2}\big|_{C_{\mu\nu}} \det \left( - \partial^2 \right)^{1/2}\big|_{F_{\mu\nu}} {\rm e}^{- \int d^4x \left( {\cal L}_{\text{\tiny $1^{+}$}} \right)} \\\nonumber
 = & \,\, {\rm const} \, .
\end{align}
Thus, the sector of transverse $2$-forms turns out to be non-physical, as it yields no contribution to the partition function.

Let us now focus on the sector of transverse vectors, i.e., the $1^-$. In this case the gauge invariant field variable is provided by Eq.\ \eqref{eq:gaugeA-inv-1-}, and the Jacobian that stems from the change of variable is
\begin{equation}
 {\cal J}^\mu{}_\nu = \delta^\mu{}_\nu \partial^2 \, .
\end{equation}
Therefore, the functional determinant on the transverse vector in Eq.\ \eqref{eq:det-funct-jac} exactly provides this change of variable. Moreover, the transverse vector that appears in the York decomposition \eqref{eq:york} is a pure gauge field variable, and it does not enter the flat-space Lagrangian Eq.\ \eqref{eq:flat-sp-lagr-1-}, but it does contribute to the functional measure, see Eq.\ \eqref{eq:det-funct-jac-metric} and \cite{Mazur:1990ak}. This allows us to write the partition function in this spin-parity sector as
\begin{align}\label{eq:part-funct-1-}
 {\cal Z_{\text{\tiny $1^-$}}} = & \det \left( - \partial^2 \right)^{1/2}\big|_{\xi_{\mu}} \int {\cal D} \zeta_{\mu} {\cal D} \tau_{\mu} \det \left( \partial^4 \right)^{1/2}\big|_{\zeta_{\mu}} \, {\rm e}^{- \int d^4x \left( {\cal L}_{\text{\tiny $1^{-}$}} \right)}\\\nonumber
 = & \det \left( - \partial^2 \right)^{1/2}\big|_{VT} \det \left( - \partial^2 \right)^{-1/2}\big|_{VT} \, .
\end{align}
The final line tells us that there is one physical transverse vector, and that we have no ghosts.

Now we turn to the $2^+$ sector. Here we have to take into account the Jacobian factor in Eq.\ \eqref{eq:det-funct-jac} that pertains to $S_{\mu\nu}$, which results in
\begin{align}\label{eq:part-funct-2+}
 {\cal Z_{\text{\tiny $2^+$}}} = & \int {\cal D} S_{\mu\nu} {\cal D} h_{\mu\nu} \det \left( - \partial^2 \right)^{1/2}\big|_{S_{\mu\nu}} \, {\rm e}^{- \int d^4x \left( {\cal L}_{\text{\tiny $2^{+}$}} \right)}\\\nonumber
 = & \det \left( - \partial^2  \right)^{-1/2}\big|_{STT} \, \det \left(  - \partial^2 \right)^{-1/2}\big|_{STT} \, \det \left( \left(\alpha\zeta^2+3\xi^2 \right) \partial^2 + 8 \varkappa^{-2} \xi^2  \right)^{-1/2}\bigg|_{STT} \, .
\end{align}
First of all, we notice that to have non-negative eigenvalues of the Laplacian we must have that
\begin{equation}
 \alpha \zeta^2 + 3 \xi^2 < 0 \, .
\end{equation}
On the other hand, the mass term already has the correct sign, and the actual flat-space mass-squared is obtained by normalizing the coefficient of the Laplacian, i.e.,
\begin{equation}
 m^2_{\text{\tiny $2^{+}$}} = \frac{8 \varkappa^{-2} \zeta^2}{| \alpha \zeta^2 + 3 \xi^2 |} \, .
\end{equation}

Finally, we analyze the behavior of the scalar sector. In this case we have non-trivial Jacobians both from the torsion and metric measure, and we get
\begin{align}\label{eq:part-funct-0+}
 {\cal Z_{\text{\tiny $0^+$}}} = & \int {\cal D} h {\cal D} \sigma {\cal D} \phi \det \left( - \partial^2 \right)^{1/2}\big|_{\phi} \det \left( - \partial^2  \right)^{1/2}\big|_{\sigma} \, {\rm e}^{- \int d^4x \left( {\cal L}_{\text{\tiny $0^{+}$}} \right)}\\\nonumber
 = & \det \left( - \partial^2 \right)^{-1/2}\big|_{{\rm scalar}}  \, \det \left( - \left( 4 \zeta^2 (\alpha - 3 \beta) + 3 \xi^2 \right) \partial^2 - 4 \varkappa^{-2} \zeta^2 \right)^{-1/2}\bigg|_{\rm scalar} \, .
\end{align}
The condition for the positive semi-definiteness of the spectrum of the Laplacian yields
\begin{equation}
 \beta > \frac{\alpha}{3} + \frac{\zeta^2}{4 \xi^2}
\end{equation}
Conversely, the sign of the mass term in Eq.\ \eqref{eq:part-funct-0+} is the wrong one, thus we have a tachyonic scalar mode. The actual value of this negative mass terms is
\begin{equation}
 m^2_{\text{\tiny $0^{+}$}} = - \frac{4 \varkappa^2 \zeta^2}{4 \zeta^2 (\alpha - 3 \beta) + 3 \xi^2} \, .
\end{equation}

In GR the presence of only two physical massless helicity modes is found by taking into account Eq.s\ \eqref{eq:part-funct-2+} and \eqref{eq:part-funct-1-}, whose appropriate limit yields the following partition function
\begin{align}\label{eq:part-funct-GR}
{\cal Z_{\text{\tiny GR}}} =  \det \left( - \partial^2 \right)^{1/2}\big|_{{\rm VT}}  \, \det \left( - \partial^2 \right)^{-1/2}\big|_{{\rm STT}} \, .
\end{align}
Transverse vectors have $3$ degrees of freedom, while symmetric transverse traceless tensors have $5$. Thus, the correct number of the degrees of freedom may be found by evaluating the zeroth Seeley-DeWitt coefficient in the heat kernel approach (see, e.g., Chapter $5$ in \cite{Percacci:2017fkn}). However, in this case we do not have such kind of cancellations among different spin-parity sectors. Thence, we cannot derive conclusive statements regarding the propagation of the degrees of freedom, apart from the absence of ghost instabilities.

While the present analysis was limited to the simplified scenario of a flat-space background, some of its outcomes are valid in more general cases. Indeed, the presence of absence of ghosts in theory is a consequence of signs of the kinetic terms only, i.e., the sole principal part of the action of a two-derivative theory is responsible for generating ghostly instabilities. Therefore, we can state that the present theory is ghost-free even of generic backgrounds.

\subsection{Some words on the phenomenological implications}	
	
The tachyonic instability found in the $0^+$ scalar sector surely hinders the phenomenological application of the present theory. However, one may hope that this unwanted behavior may due to the oversimplifying assumption of a flat-space background. In this regard, the first non-trivial generalization that we can take into account is given by having a maximally symmetric background Riemannian structure, and still assuming a vanishing torsion tensor. In this case we find that the absence of tachyons in the scalar sector requires that
\begin{equation}
	R<0 \, , \qquad \beta> \frac{\alpha}{3} + \frac{\zeta^2}{\xi^2} - \frac{1}{R \varkappa^2} \, .
\end{equation}
Therefore, the background space must be the Euclidean AdS, i.e., a hyperboloid. However, the correct identification of the propagating degrees of freedom on a general background is be done in a rigorous manner by performing a Hamiltonian analysis of the theory in Minkowski signature. In this case a different physical phenomenon may arise, which is the spontaneous breaking of the gauge invariance that we have derived. In this case the ``eaten" particle would be the two massive excitations of the metric tensor, and one could wonder if the antisymmetric part of the torsion may generate a non-trivial background configuration providing a novel kind of vacuum state which is compatible with the cosmological principle. In such a case the masses could also receive non-trivial non-local contributions similar \emph{à la} Coleman-Weinberg \cite{Weinberg:1973am}. In this sense the purely torsional part of the theory is manifestly scale invariant, and it will thus be expected to give rise to dimensional transmutation once the one-loop effects are properly taken into account. Thence, it would be compelling to analyze the physical implications of this quantum phenomenon on the nature of the propagating degrees of freedom. Nevertheless, these topics lie beyond the scope of the present paper, and they shall not be discussed henceforth.

A partially unrelated point that needs to be addressed about this new model is represented by singling out its phenomenological implications regarding the interactions of the torsion with Standard Model (SM) fields, and more generally by identifying which of its predictions that can be tested in the near future. Since the gauge invariance of the action precludes any interaction between the torsion and the Standard Model fields, any amplitude of, say, torsion particles to SM ones necessarily requires the presence of intermediate off-shell metric perturbations, and below the mass threshold the latter will be mostly given by gravitons. Thus, the phenomenological fingerprint of this theory is qualitatively represented by an enhancement in the particle production in the early Universe, whose leading contribution will be given by the decay of torsion particles into gravitons due to the linear term in the action, Eq.\ \eqref{eq:inv-act-o1}. Moreover, this term will also be responsible for a modified power spectra of primordial gravitational waves and scalar perturbations, whose cosmological consequences can provide phenomenological bounds on the couplings of the model.

We now compare the present model to those that have been proposed in the literature throughout the years. Due to the very complicated nature of the most general torsion action, much of the effort in putting forward torsionful completions of General Relativity was focused on the so-called Poincaré gauge theories \cite{Hehl:1976kj,Hehl:1978yt,Hehl:1994ue,Gronwald:1995em}. In this sense, the previous analyses \cite{Neville:1979rb,Sezgin:1979zf,Sezgin:1981xs} focused mostly on constraining the parameter space by requiring the absence of ghosts and tachyons, and they generally include three massive propagating spin-parity sectors beside the graviton. However, due to the \emph{ad hoc} requirements that are assumed, these models are non-renormalizable \cite{Melichev:2023lwj}, since the radiative corrections generate counterterms whose tensor structures radically differs from those that appear in the starting action. Indeed, up to now these is no proposed torsion model which is both devoid of instabilities and renormalizable. A more complete formalism for analyzing the stability in general metric-affine theories was developed in \cite{Baldazzi:2021kaf}; however, due to the enormous number of the free parameters involved the full stability analysis was only performed in some simplified toy-models. Therefore, the present model stands on a completely different footing with respect to those that have been brought forward previously, since it focuses on the presence of an off-shell gauge symmetry to both gain predictive power by constraining the parameter space and to stabilize the renormalization-group flow on a renormalizable trajectory. While the first aim has been certainly achieved, the second one is far from being settled. With this new perspective in mind the stability and particle content of the theory are to be thought as predictions of a given model, and they serve as a criterion to discriminate between ``healthy" and ``sick" gauge symmetries.

\subsection{Remarks on the $1$-loop fluctuations on a generic background}\label{subsect:quantum-remarks}

Working out the divergent part of the effective action due to torsion fluctuations is generally a complicated task. As a matter of fact, this calculation is far from being doable even in the manifestly covariant formalism provided by the heat kernel technique. The reason is that the torsion has three algebraically irreducible covariant divergences, that give rise to three different non-minimal terms in the principal part of the Hessian. These non-minimal terms are usually counterbalanced by the gauge-fixing in Yang-Mills gauge theories and gravity itself. Furthermore, the torsion-independent part of the Lagrangian also involves a fourth-order operator acting on the metric perturbations. Thus, one has to perform perturbation theory in order to appropriately take into account the off-diagonal mixing of the torsion and metric perturbations. When it is possible to cancel out all the non-minimal terms through an appropriate choice of the gauge-fixing, the $1$-loop computation can be done by employing the known asymptotic expansion of the heat kernel \cite{Obukhov:1983mm,Barvinsky:1985an}.

There are paths one can follow in order to circumvent this problematic issue. One can either choose a specific Lagrangian that involves only one algebraically irreducible non-minimal term, or opt for performing a covariant spin-parity decomposition to get rid of the non-minimal terms. The first route was taken in \cite{Melichev:2023lwj}, while the second was attempted in \cite{Martini:2023apm}. We stress that the second possibility comes with a huge price, i.e., having to deal with higher-order operators.

The conceptual problem of choosing a specific Lagrangian is that one does not generally expect it to be stable along the Renormalization Group (RG) flow, unless there is some underlying symmetry that has been used to build it in the first place. In fact, such a non-closure of the RG flow is exactly what has been found in \cite{Melichev:2023lwj}. Therefore, our quest for a further gauge symmetry is also meant to select a stable trajectory in the RG flow of torsionful theories.

In what follows, we will succinctly show that our gauge principle Eq.\ \eqref{eq:deltaG} makes it possible to obtain a minimal differential operator from the torsion Hessian of Eq.\ \eqref{eq:full-action}. Thus, a heat kernel computation of the beta functions can be undertaken in this case. First of all, let us split the torsion into its completely antisymmetric and hook antisymmetric parts, i.e.,
\begin{equation}\label{eq:alg-decomp-t-a}
T^\rho{}_{\mu\nu} = t^\rho{}_{\mu\nu} + a^\rho{}_{\mu\nu} \, .
\end{equation}
where
\begin{eqnarray}
 t_{[\rho\mu\nu]} = 0 \, , && \qquad a_{\rho\mu\nu} = a_{[\rho\mu\nu]} \, .
\end{eqnarray}
For notational convenience we employ the following shorthand for the gradient and divergence of the two components of the torsion, i.e., 
\begin{eqnarray}
\left({\rm grad}\, t\right)_\lambda{}^\rho{}_{\mu\nu} = \nabla_\lambda t^\rho{}_{\mu\nu} \, , && \qquad \left({\rm div}_1\, t\right)_{\mu\nu} = \nabla_\rho t^\rho{}_{\mu\nu} \, ,
\end{eqnarray}
and similarly for $a^\rho{}_{\mu\nu}$. Accordingly, by substituting the decomposition Eq.\ \eqref{eq:alg-decomp-t-a} into the kinetic terms that we read off from Eq.\ \eqref{eq:inv-act-o2}, we obtain
\begin{align}\label{eq:action-kin-red}
 S_{\rm kin} [g,T] = \int \sqrt{g} & \left[ \frac{1}{6} ({\rm grad} \, t)^2 + \frac{8}{3} ({\rm grad} \, a)^2 - \frac{1}{6} ({\rm div}_1 \, t)^2 - \frac{8}{3} ({\rm div}_1 \, a)^2  \right.\\\nonumber
 & \,\,\, \left. - \frac{4}{3} ({\rm div}_1 \, t) \cdot ({\rm div}_1 \, a) \right] \, ,
\end{align}
where the dot stands for contraction of the indices.

The gauge-fixing function of the transformation \eqref{eq:deltaG} is a two-form. Then, up to the addition of terms that will depend on the background torsion, it will then take the following form
\begin{equation}
 f_{\mu\nu} = ({\rm div}_1 \, t)_{\mu\nu} + \beta ({\rm div}_1 \,  H)_{\mu\nu} \, .
\end{equation}
Accordingly, its contribution to the action will be given by
\begin{equation}
 S_{\rm gf} [g,T] = \frac{1}{\alpha} \int \sqrt{g} \, f_{\mu\nu} f^{\mu\nu} \, . 
\end{equation}
By choosing $\alpha=\frac{1}{6}$ and $\beta= 8$ we cancel out exactly the non-minimal terms of Eq.\ \eqref{eq:action-kin-red}. Therefore, the sum of the kinetic action plus the gauge-fixing yields the subsequent minimal kinetic terms for the two components of the torsion
\begin{equation}
 S_{\rm kin} [g,T] + S_{\rm gf} [g,T] = \int \sqrt{g}  \left[ \frac{1}{6} ({\rm grad} \, t)^2 + \frac{8}{3} ({\rm grad} \, a)^2 \right] \, .
\end{equation}
The Hessian of the torsion fluctuations is thus devoid of non-minimal terms, yet these arise in the ghost Lagrangian. Indeed, the operator in the ghost action is found by varying the gauge-fixing $2$-form $f_{\mu\nu}$ w.r.t.\ \eqref{eq:deltaG}, which yields the following operator
\begin{equation}
 (\Delta A)_{\mu\nu} = \frac{16}{3} \left( \square A_{\mu\nu} - \frac{1}{4} \nabla_\lambda \nabla_\mu A^\lambda{}_\nu + \frac{1}{4} \nabla_\lambda \nabla_\nu A^\lambda{}_\mu \right) \, .
\end{equation}
Therefore, the complicated part of the calculation is evaluating the ghost contribution. Nevertheless, it is always possible to choose the ghost action such it present no derivative interactions, which drastically simplifies the evaluation.

Since the full action functional \eqref{eq:full-action} contains derivative interactions, the $1$-loop integration of the torsion fluctuations may be carried out exploiting the results of \cite{Obukhov:1983mm}. Furthermore, to take into account the metric fluctuations we would need the results for minimal fourth-order operators that have been derived in \cite{Barvinsky:1985an}. Finally, we would have to perform perturbation theory to take into account the off-diagonal mixing of the two types of quantum disturbances.

In summary, our new gauge principle \eqref{eq:deltaG} makes it possible to choose a minimal gauge for torsion fluctuations. In spite of this, the ghost action does display non-minimal modes, whereas the derivative interactions are absent if the gauge-fixing action is appropriately chosen. We defer this highly interesting yet lengthy analysis to future work.

\section{Conclusions and outlooks}\label{sect:conclusions}

In this paper we have performed for the first time a systematic analysis of the possible gauge invariances that give rise to closed algebras and can be consistently imposed in torsionful theories of gravity. The gauge transformations that we have focused on are affine in the torsion, they are parameterized by even-parity tensors and involve only one covariant derivative of these tensors. Therefore, one possible extension of the present work may be to consider transformations that are parameterized by odd-symmetry tensors, or that involve more covariant derivatives. Nevertheless, we remark that our hypotheses regarding the ansatze of the transformations are in complete analogy with those that one usually encounters in Yang-Mills \cite{Weinberg:1996kr}, General Relativity \cite{Percacci:2017fkn} and higher-spin gauge theories \cite{Ponomarev:2022vjb}. Another possible future development of the present work is to apply the same technique to general metric-affine theories, i.e., those that have a non-trivial non-metricity.

The most interesting result that we have obtained is that it is possible to impose a non-abelian gauge structure \eqref{eq:algebra}, that may be interpreted as a holonomic twin of the local Lorentz algebra. By imposing the invariance under such a transformation (see Eq.\ \eqref{eq:deltaG}) we were able to derive an invariant action \eqref{eq:full-action}, which has $4$ arbitrary couplings instead of the $94$ of the most general power-counting renormalizable action. Furthermore, the flat-space content of the theory is devoid of ghost instabilities, whereas it does present a tachyonic scalar mode. However, this mode has a healthy mass term on a maximally symmetric background with negative curvature. Therefore, a more detailed investigation of the physical properties of this model is required in order to make precise claims regarding the nature of the mass terms.

Our analysis in Sect.\ \ref{sect:non-lin} has also yielded an abelian gauge structure, which is parameterized by a scalar. The infinitesimal transformation is a non-trivial generalization of the one that has already been studied in the literature \cite{Shapiro:2001rz,Sauro:2022chz,Sauro:2022hoh,Paci:2024ohq}, the difference being that the known one only acts on the torsion vector. In this work we have not worked out any phenomenological aspect of the associated invariant action, therefore this will certainly be a possible future development.

As we have explained both in the Introduction \ref{sect:intro} and in Sect.\ \ref{sect:gauge-inv-dof}, we have chosen to rely on the flat-space path integral to check the stability of our model. Indeed, it was argued in \cite{Mazur:1990ak} that this method circumvents the appearance of some unphysical states that are found using spin projectors. Nevertheless, the final word can only be said by performing a $3+1$ decomposition of the model, followed by a transverse decomposition of the resulting tensors. Therefore, a non-trivial and necessary development of the present paper is a thorough analysis of the $3+1$ decomposition in metric-affine theories. Moreover, the $3+1$ decomposition is also necessary for an exact counting of the propagating degrees of freedom. As a matter of fact, this study would also allow to us systematically analyze which types of “hairs” can appear in metric-affine theories.

In the last part of the paper \ref{subsect:quantum-remarks} we have showed that imposing the gauge invariance \eqref{eq:deltaG} makes it possible to choose a minimal gauge for torsion fluctuations. Therefore, a natural and necessary advancement of the present work would be a $1$-loop integration of the quantum fluctuations of the action \eqref{eq:full-action}. The complicated part of this program lies in the fact that the ghost action displays non-minimal kinetic terms, which may be accounted for following the method outlined in \cite{Barvinsky:1985an}. We remark that the metric sector of the theory has a higher-derivative operator, thus one would face this kind of complication too. Having said that, we stress that performing the calculation of the divergent part of the effective action is crucial for a better understanding of the theory. Indeed, this would firstly allow to check if the gauge invariance \eqref{eq:deltaG} is non-anomalous, and secondly it would provide the necessary ingredients to analyze the RG flow of the theory and the possible presence of fixed-points of the RG equation \cite{Wetterich:1992yh}. This last point is of paramount importance, for the UV consistency of the model lingers on a the presence of an asymptotically safe fixed-point.

Finally, we mention some further points regarding the non-abelian gauge structure \eqref{eq:full-gauge-algebra} that we have skipped in our preliminary analysis, but which need to be expanded in future works. First of all, a global interpretation of the gauge algebra is unclear, and this would definitely be of paramount importance for understanding the non-perturbative features of the invariant action \eqref{eq:full-action}. Secondly, it would be compelling to uncover the physical features of the Noether charges that are associated with the Noether identity \eqref{eq:noetherG}, and this point immediately brings about the necessity to build gauge-invariant asymptotic states \cite{Grassi:2024vkb}. Eventually, the construction of such states would permit to set the stage for the study of symmetry breaking phenomena, as well as phenomenological applications of the model to the inflationary cosmology.

\section{Acknowledgements}\label{sect:ack}

The author wishes to thank Omar Zanusso for useful comments on the draft and for profitable discussions.

\appendix


\section{Closure of the algebra: intermediate steps}\label{sect:app-closed-algebra}

Combining Eq.s\ \eqref{eq:ansatz-deltaA} and \eqref{eq:ansatz-deltaPi} we have the most general affine transformation parametrized by a two-form $A_1{}_{\mu\nu}$, i.e.,
\begin{equation}\label{eq:ansatz-deltaAPi}
	\begin{split}
			& \delta^G{}_{A_1} T^\rho{}_{\mu\nu} = (-2 c_2 + b_2 \theta) A_1{}^{\rho \alpha } T_{\alpha \mu \nu } + (- c_2 -  b_2 \theta) A_1{}_{\nu }{}^{\alpha } T_{\alpha \mu }{}^{\rho } + (c_2 + b_2 \theta) A_1{}_{\mu }{}^{\alpha } T_{\alpha \nu }{}^{\rho }\\
			& + (- c_3 + b_3 \theta) A_1{}_{\nu }{}^{\rho } T^{\alpha }{}_{\mu \alpha } + (c_3 -  b_3 \theta) A_1{}_{\mu }{}^{\rho } T^{\alpha }{}_{\nu \alpha } + (2 c_3 + b_3 \theta) A_1{}_{\mu \nu } T^{\alpha \rho }{}_{\alpha } \\
			& + (c_1 -  b_1 \theta) A_1{}^{\rho \alpha } T_{\mu \nu \alpha } + b_1 \theta A_1{}_{\nu }{}^{\alpha } T_{\mu }{}^{\rho }{}_{\alpha } + (- c_1 + b_1 \theta) A_1{}^{\rho \alpha } T_{\nu \mu \alpha } \\
			& -  b_1 \theta A_1{}_{\mu }{}^{\alpha } T_{\nu }{}^{\rho }{}_{\alpha } + (- c_1 -  b_1 \theta) A_1{}_{\nu }{}^{\alpha } T^{\rho }{}_{\mu \alpha } + (c_1 + b_1 \theta) A_1{}_{\mu }{}^{\alpha } T^{\rho }{}_{\nu \alpha } \\
			& + (-1 + \theta) \nabla_{\mu }A_1{}_{\nu }{}^{\rho } + (1 -  \theta) \nabla_{\nu }A_1{}_{\mu }{}^{\rho } + (2 + \theta) \nabla^{\rho }A_1{}_{\mu \nu } \, .
		\end{split}
	\end{equation}
Here $\theta$ is the ``angle" which sets the relative weights of the transformations \eqref{eq:ansatz-deltaA} and \eqref{eq:ansatz-deltaPi}. Using this explicit expression we compute the commutator between two transformations, obtaining
{\footnotesize
\begin{align}
			& \left[ \delta^G_{A_1} , \delta^G_{A_2} \right] T^\rho{}_{\mu\nu} = (-2 c_1 + 3 c_2) (c_3 -  b_3 \theta) A_1{}_{\nu }{}^{\rho } A_2{}^{\alpha \sigma} T_{\alpha \mu \sigma} \\\nonumber
			& + \bigl(- c_1^2 + c_2^2 + 3 b_1 c_1 \theta -  (b_1 - 2 b_2) c_2 \theta + (-2 b_1^2 -  b_1 b_2 + b_2^2) \theta^2\bigr) A_1{}^{\rho \alpha } A_2{}_{\nu }{}^{\sigma} T_{\alpha \mu \sigma}\\\nonumber
			& + (2 c_1 - 3 c_2) (c_3 -  b_3 \theta) A_1{}^{\alpha \sigma} A_2{}_{\nu }{}^{\rho } T_{\alpha \mu \sigma} + \bigl(- c_1^2 + 2 c_2^2 - 3 b_2 c_1 \theta + (-2 b_1 + b_2) c_2 \theta + (2 b_1^2 + b_1 b_2 -  b_2^2) \theta^2\bigr) A_1{}_{\nu }{}^{\alpha } A_2{}^{\rho \sigma} T_{\alpha \mu \sigma}\\\nonumber
			& + (2 c_1 - 3 c_2) (c_3 -  b_3 \theta) A_1{}_{\mu }{}^{\rho } A_2{}^{\alpha \sigma} T_{\alpha \nu \sigma} + \bigl(c_1^2 -  c_2^2 - 3 b_1 c_1 \theta + (b_1 - 2 b_2) c_2 \theta + (2 b_1^2 + b_1 b_2 -  b_2^2) \theta^2\bigr) A_1{}^{\rho \alpha } A_2{}_{\mu }{}^{\sigma} T_{\alpha \nu \sigma} \\\nonumber
			& + (-2 c_1 + 3 c_2) (c_3 -  b_3 \theta) A_1{}^{\alpha \sigma} A_2{}_{\mu }{}^{\rho } T_{\alpha \nu \sigma} + \bigl(c_1^2 - 2 c_2^2 + 3 b_2 c_1 \theta + (2 b_1 -  b_2) c_2 \theta + (-2 b_1^2 -  b_1 b_2 + b_2^2) \theta^2\bigr) A_1{}_{\mu }{}^{\alpha } A_2{}^{\rho \sigma} T_{\alpha \nu \sigma}\\\nonumber
			& + (2 c_1 - 3 c_2) (2 c_3 + b_3 \theta) A_1{}_{\mu \nu } A_2{}^{\alpha \sigma} T_{\alpha }{}^{\rho }{}_{\sigma} + \bigl(c_2^2 + 2 (b_1 + b_2) c_2 \theta + (-2 b_1^2 -  b_1 b_2 + b_2^2) \theta^2\bigr) A_1{}_{\nu }{}^{\alpha } A_2{}_{\mu }{}^{\sigma} T_{\alpha }{}^{\rho }{}_{\sigma} \\\nonumber
			& + (-2 c_1 + 3 c_2) (2 c_3 + b_3 \theta) A_1{}^{\alpha \sigma} A_2{}_{\mu \nu } T_{\alpha }{}^{\rho }{}_{\sigma} + \bigl(- c_2^2 - 2 (b_1 + b_2) c_2 \theta + (2 b_1^2 + b_1 b_2 -  b_2^2) \theta^2\bigr) A_1{}_{\mu }{}^{\alpha } A_2{}_{\nu }{}^{\sigma} T_{\alpha }{}^{\rho }{}_{\sigma} \\\nonumber
			& + \bigl(c_1^2 - 2 c_2^2 + 3 b_2 c_1 \theta + (2 b_1 -  b_2) c_2 \theta + (-2 b_1^2 -  b_1 b_2 + b_2^2) \theta^2\bigr) A_1{}^{\rho \alpha } A_2{}_{\nu }{}^{\sigma} T_{\sigma \mu \alpha } \\\nonumber
			& + \bigl(c_1^2 - c_2^2 - 3 b_1 c_1 \theta + (b_1 - 2 b_2) c_2 \theta + (2 b_1^2 + b_1 b_2 - b_2^2) \theta^2\bigr) A_1{}_{\nu }{}^{\alpha } A_2{}^{\rho \sigma} T_{\sigma \mu \alpha } \\\nonumber
			& + \bigl(-4 c_2^2 - 2 c_1 (c_2 + b_2 \theta) -  b_2 \theta (c_4 - 2 b_1 \theta + b_2 \theta) + 2 c_2 (c_4 + b_1 \theta + 2 b_2 \theta)\bigr) A_1{}^{\rho \alpha } A_2{}_{\alpha }{}^{\sigma} T_{\sigma \mu \nu } \\\nonumber
			& + \bigl(4 c_2^2 + 2 c_1 (c_2 + b_2 \theta) + b_2 \theta (c_4 - 2 b_1 \theta + b_2 \theta) - 2 c_2 (c_4 + b_1 \theta + 2 b_2 \theta)\bigr) A_1{}^{\alpha \sigma} A_2{}^{\rho }{}_{\alpha } T_{\sigma \mu \nu } \\\nonumber
			& + (c_2 + b_2 \theta) (- c_1 - 2 c_2 + c_4 - 2 b_1 \theta + b_2 \theta) A_1{}_{\nu}{}^{\alpha } A_2{}_{\alpha }{}^{\sigma} T_{\sigma \mu }{}^{\rho } + (c_1 + 2 c_2 -  c_4 + 2 b_1 \theta -  b_2 \theta) (c_2 + b_2 \theta) A_1{}^{\alpha \sigma} A_2{}_{\nu \alpha } T_{\sigma \mu }{}^{\rho } \\\nonumber
			& + \bigl(- c_1^2 + 2 c_2^2 - 3 b_2 c_1 \theta + (-2 b_1 + b_2) c_2 \theta + (2 b_1^2 + b_1 b_2 -  b_2^2) \theta^2\bigr) A_1{}^{\rho \alpha } A_2{}_{\mu }{}^{\sigma} T_{\sigma \nu \alpha } \\\nonumber
			& + \bigl(- c_1^2 + c_2^2 + 3 b_1 c_1 \theta -  (b_1 - 2 b_2) c_2 \theta + (-2 b_1^2 -  b_1 b_2 + b_2^2) \theta^2\bigr) A_1{}_{\mu }{}^{\alpha } A_2{}^{\rho \sigma} T_{\sigma \nu \alpha }\\\nonumber
			& + (c_1 + 2 c_2 -  c_4 + 2 b_1 \theta - b_2 \theta) (c_2 + b_2 \theta) A_1{}_{\mu }{}^{\alpha } A_2{}_{\alpha }{}^{\sigma} T_{\sigma \nu }{}^{\rho } + (c_2 + b_2 \theta) (- c_1 - 2 c_2 + c_4 - 2 b_1 \theta + b_2 \theta) A_1{}^{\alpha \sigma} A_2{}_{\mu \alpha } T_{\sigma \nu }{}^{\rho } \\\nonumber
			& + \bigl(c_2^2 + 2 (b_1 + b_2) c_2 \theta + (-2 b_1^2 -  b_1 b_2 + b_2^2) \theta^2\bigr) A_1{}_{\nu }{}^{\alpha } A_2{}_{\mu }{}^{\sigma} T_{\sigma}{}^{\rho }{}_{\alpha } \\\nonumber
			& + \bigl(- c_2^2 - 2 (b_1 + b_2) c_2 \theta + (2 b_1^2 + b_1 b_2 -  b_2^2) \theta^2\bigr) A_1{}_{\mu }{}^{\alpha } A_2{}_{\nu }{}^{\sigma} T_{\sigma}{}^{\rho }{}_{\alpha } \\\nonumber
			& + \bigl(3 c_3^2 + (2 b_1 + 2 b_2 - 3 b_3) c_3 \theta + (-2 b_1 + b_2) b_3 \theta^2 + c_2 (2 c_3 + b_3 \theta)\bigr) A_1{}_{\nu }{}^{\rho } A_2{}_{\mu }{}^{\alpha } T^{\sigma}{}_{\alpha \sigma}\\\nonumber
			& + \bigl(6 c_3^2 + (-2 b_1 - 2 b_2 + 3 b_3) c_3 \theta + b_3 \theta (-3 c_1 + 2 b_1 \theta -  b_2 \theta) + 2 c_2 (2 c_3 + b_3 \theta)\bigr) A_1{}^{\rho \alpha } A_2{}_{\mu \nu } T^{\sigma}{}_{\alpha \sigma} \\\nonumber
			& + \bigl(3 c_3^2 + (2 b_1 + 2 b_2 - 3 b_3) c_3 \theta + (-2 b_1 + b_2) b_3 \theta^2 + c_2 (2 c_3 + b_3 \theta)\bigr) A_1{}_{\nu }{}^{\alpha } A_2{}_{\mu }{}^{\rho } T^{\sigma}{}_{\alpha \sigma}\\\nonumber
			& + \bigl(-3 c_3^2 + (-2 b_1 - 2 b_2 + 3 b_3) c_3 \theta + (2 b_1 -  b_2) b_3 \theta^2 - c_2 (2 c_3 + b_3 \theta)\bigr) A_1{}_{\mu }{}^{\rho } A_2{}_{\nu }{}^{\alpha } T^{\sigma}{}_{\alpha \sigma} \\\nonumber
			& + \bigl(-3 c_3^2 + (-2 b_1 - 2 b_2 + 3 b_3) c_3 \theta + (2 b_1 -  b_2) b_3 \theta^2 -  c_2 (2 c_3 + b_3 \theta)\bigr) A_1{}_{\mu }{}^{\alpha } A_2{}_{\nu }{}^{\rho } T^{\sigma}{}_{\alpha \sigma} \\\nonumber
			& + \bigl(-6 c_3^2 + (2 b_1 + 2 b_2 - 3 b_3) c_3 \theta + b_3 \theta (3 c_1 - 2 b_1 \theta + b_2 \theta) - 2 c_2 (2 c_3 + b_3 \theta)\bigr) A_1{}_{\mu \nu } A_2{}^{\rho \alpha } T^{\sigma}{}_{\alpha \sigma} \\\nonumber
			& + \bigl(- c_2 c_3 + c_3 c_4 + b_3 c_2 \theta + 2 b_1 c_3 \theta + 2 b_2 c_3 \theta -  b_3 c_4 \theta + 4 b_1 b_3 \theta^2 - 2 b_2 b_3 \theta^2 -  c_1 (2 c_3 + b_3 \theta)\bigr) A_1{}^{\rho \alpha } A_2{}_{\nu \alpha } T^{\sigma}{}_{\mu \sigma}\\\nonumber
			& + \Bigl(- c_3 c_4 + b_3 c_4 \theta + c_2 (c_3 -  b_3 \theta) + c_1 (2 c_3 + b_3 \theta) - 2 \theta \bigl(b_2 (c_3 -  b_3 \theta) + b_1 (c_3 + 2 b_3 \theta)\bigr)\Bigr) A_1{}_{\nu }{}^{\alpha } A_2{}^{\rho }{}_{\alpha } T^{\sigma}{}_{\mu \sigma} \\\nonumber
			& + \Bigl(- c_3 c_4 + b_3 c_4 \theta + c_2 (c_3 -  b_3 \theta) + c_1 (2 c_3 + b_3 \theta) - 2 \theta \bigl(b_2 (c_3 -  b_3 \theta) + b_1 (c_3 + 2 b_3 \theta)\bigr)\Bigr) A_1{}^{\rho \alpha } A_2{}_{\mu \alpha } T^{\sigma}{}_{\nu \sigma}\\\nonumber
			& + \Bigl(c_3 c_4 -  b_3 c_4 \theta + c_2 (- c_3 + b_3 \theta) -  c_1 (2 c_3 + b_3 \theta) + 2 \theta \bigl(b_2 (c_3 -  b_3 \theta) + b_1 (c_3 + 2 b_3 \theta)\bigr)\Bigr) A_1{}_{\mu }{}^{\alpha } A_2{}^{\rho }{}_{\alpha } T^{\sigma}{}_{\nu \sigma} \\\nonumber
			& + \Bigl(- c_4 (2 c_3 + b_3 \theta) + 2 \bigl(c_2 (c_3 -  b_3 \theta) + c_1 (2 c_3 + b_3 \theta) + \theta (b_1 c_3 + b_2 c_3 + 2 b_1 b_3 \theta - b_2 b_3 \theta)\bigr)\Bigr) A_1{}_{\nu }{}^{\alpha } A_2{}_{\mu \alpha } T^{\sigma \rho }{}_{\sigma}\\\nonumber
			& + \Bigl(c_4 (2 c_3 + b_3 \theta) - 2 \bigl(c_2 (c_3 -  b_3 \theta) + c_1 (2 c_3 + b_3 \theta) + \theta (b_1 c_3 + b_2 c_3 + 2 b_1 b_3 \theta -  b_2 b_3 \theta)\bigr)\Bigr) A_1{}_{\mu }{}^{\alpha } A_2{}_{\nu \alpha } T^{\sigma \rho }{}_{\sigma} \\\nonumber
			& + c_1 (c_3 -  b_3 \theta) A_1{}_{\nu }{}^{\rho } A_2{}^{\alpha \sigma} T_{\mu \alpha \sigma} + (c_1^2 - 3 b_1 c_1 \theta + 3 b_1 c_2 \theta) A_1{}^{\rho \alpha } A_2{}_{\nu }{}^{\sigma} T_{\mu \alpha \sigma} + c_1 (- c_3 + b_3 \theta) A_1{}^{\alpha \sigma} A_2{}_{\nu }{}^{\rho } T_{\mu \alpha \sigma}\\\nonumber
			& + (c_1^2 - 3 b_1 c_1 \theta + 3 b_1 c_2 \theta) A_1{}_{\nu }{}^{\alpha } A_2{}^{\rho \sigma} T_{\mu \alpha \sigma} + (c_1 -  b_1 \theta) (c_1 + 2 c_2 -  c_4 + 2 b_1 \theta -  b_2 \theta) A_1{}^{\rho \alpha } A_2{}_{\alpha }{}^{\sigma} T_{\mu \nu \sigma} \\\nonumber
			& -  (c_1 -  b_1 \theta) (c_1 + 2 c_2 -  c_4 + 2 b_1 \theta -  b_2 \theta) A_1{}^{\alpha \sigma} A_2{}^{\rho }{}_{\alpha } T_{\mu \nu \sigma} \\\nonumber
			& + \Bigl(b_1 \theta (- c_2 -  c_4 + 2 b_1 \theta -  b_2 \theta) + c_1 \bigl(c_2 + (2 b_1 + b_2) \theta \bigr)\Bigr) A_1{}_{\nu }{}^{\alpha } A_2{}_{\alpha }{}^{\sigma} T_{\mu }{}^{\rho }{}_{\sigma} \\\nonumber
			& + \Bigl(b_1 \theta (c_2 + c_4 - 2 b_1 \theta + b_2 \theta) -  c_1 \bigl(c_2 + (2 b_1 + b_2) \theta \bigr)\Bigr) A_1{}^{\alpha \sigma} A_2{}_{\nu \alpha } T_{\mu }{}^{\rho }{}_{\sigma} + c_1 (- c_3 + b_3 \theta) A_1{}_{\mu }{}^{\rho } A_2{}^{\alpha \sigma} T_{\nu \alpha \sigma} \\\nonumber
			& + (- c_1^2 + 3 b_1 c_1 \theta - 3 b_1 c_2 \theta) A_1{}^{\rho \alpha } A_2{}_{\mu }{}^{\sigma} T_{\nu \alpha \sigma} + c_1 (c_3 -  b_3 \theta) A_1{}^{\alpha 	\sigma} A_2{}_{\mu }{}^{\rho } T_{\nu \alpha \sigma} + (- c_1^2 + 3 b_1 c_1 \theta - 3 b_1 c_2 \theta) A_1{}_{\mu }{}^{\alpha } A_2{}^{\rho \sigma} T_{\nu \alpha \sigma}\\\nonumber
			& + (- c_1 + b_1 \theta) (c_1 + 2 c_2 -  c_4 + 2 b_1 \theta -  b_2 \theta) A_1{}^{\rho \alpha } A_2{}_{\alpha }{}^{\sigma} T_{\nu \mu \sigma}  + (c_1 -  b_1 \theta) (c_1 + 2 c_2 -  c_4 + 2 b_1 \theta -  b_2 \theta) A_1{}^{\alpha \sigma} A_2{}^{\rho }{}_{\alpha } T_{\nu \mu \sigma} \\\nonumber
			& + \Bigl(b_1 \theta (c_2 + c_4 - 2 b_1 \theta + b_2 \theta) -  c_1 \bigl(c_2 + (2 b_1 + b_2) \theta \bigr)\Bigr) A_1{}_{\mu }{}^{\alpha } A_2{}_{\alpha }{}^{\sigma} T_{\nu }{}^{\rho }{}_{\sigma} \\\nonumber
			& + \Bigl(b_1 \theta (- c_2 -  c_4 + 2 b_1 \theta -  b_2 \theta) + c_1 \bigl(c_2 + (2 b_1 + b_2) \theta \bigr)\Bigr) A_1{}^{\alpha \sigma} A_2{}_{\mu \alpha } T_{\nu }{}^{\rho }{}_{\sigma} -  c_1 (2 c_3 + b_3 \theta) A_1{}_{\mu \nu } A_2{}^{\alpha \sigma} T^{\rho }{}_{\alpha \sigma} \\\nonumber
			& + c_1 (2 c_3 + b_3 \theta) A_1{}^{\alpha \sigma} A_2{}_{\mu \nu } T^{\rho }{}_{\alpha \sigma} + \bigl(- c_1^2 + b_1 \theta (c_2 + c_4 - 2 b_1 \theta + b_2 \theta) -  c_1 (c_2 -  c_4 + 2 b_1 \theta + b_2 \theta)\bigr) A_1{}_{\nu }{}^{\alpha } A_2{}_{\alpha }{}^{\sigma} T^{\rho }{}_{\mu \sigma} \\\nonumber
			& + \Bigl(- c_4 (c_1 + b_1 \theta) + b_1 \theta (c_1 -  c_2 + 2 b_1 \theta -  b_2 \theta) + c_1 \bigl(c_1 + c_2 + (b_1 + b_2) \theta \bigr)\Bigr) A_1{}^{\alpha \sigma} A_2{}_{\nu \alpha } T^{\rho }{}_{\mu \sigma} \\\nonumber
			& + \Bigl(- c_4 (c_1 + b_1 \theta) + b_1 \theta (c_1 -  c_2 + 2 b_1 \theta -  b_2 \theta) + c_1 \bigl(c_1 + c_2 + (b_1 + b_2) \theta \bigr)\Bigr) A_1{}_{\mu }{}^{\alpha } A_2{}_{\alpha }{}^{\sigma} T^{\rho }{}_{\nu \sigma} \\\nonumber
			& + \Bigl(c_4 (c_1 + b_1 \theta) + b_1 \theta \bigl(- c_1 + c_2 + (-2 b_1 + b_2) \theta \bigr) -  c_1 \bigl(c_1 + c_2 + (b_1 + b_2) \theta \bigr)\Bigr) A_1{}^{\alpha \sigma} A_2{}_{\mu \alpha } T^{\rho }{}_{\nu \sigma} \\\nonumber
			& + (-3 c_3 + 3 b_3 \theta) A_2{}_{\nu }{}^{\rho } \nabla_{\alpha }A_1{}_{\mu }{}^{\alpha } + \bigl(2 c_1 (-1 + \theta) - 2 b_1 (-1 + \theta) \theta - 2 c_2 (2 + \theta) + b_2 \theta (2 + \theta)\bigr) A_2{}^{\rho \alpha } \nabla_{\alpha }A_1{}_{\mu \nu } \\\nonumber
			& + \bigl(c_1 (-1 + \theta) + 2 b_1 (-1 + \theta) \theta -  c_2 (2 + \theta) -  b_2 \theta (2 + \theta)\bigr) A_2{}_{\nu }{}^{\alpha } \nabla_{\alpha }A_1{}_{\mu }{}^{\rho } + 3 (c_3 -  b_3 \theta) A_2{}_{\mu }{}^{\rho } \nabla_{\alpha }A_1{}_{\nu }{}^{\alpha } \\\nonumber
			& + \bigl(c_1 -  c_1 \theta - 2 b_1 (-1 + \theta) \theta + c_2 (2 + \theta) + b_2 \theta (2 + \theta)\bigr) A_2{}_{\mu }{}^{\alpha } \nabla_{\alpha }A_1{}_{\nu }{}^{\rho } + (6 c_3 + 3 b_3 \theta) A_2{}_{\mu \nu } \nabla_{\alpha }A_1{}^{\rho \alpha } \\\nonumber
			& + 3 (c_3 -  b_3 \theta) A_1{}_{\nu }{}^{\rho } \nabla_{\alpha }A_2{}_{\mu }{}^{\alpha } + \bigl(-2 c_1 (-1 + \theta) + 2 b_1 (-1 + \theta) \theta + 2 c_2 (2 + \theta) -  b_2 \theta (2 + \theta)\bigr) A_1{}^{\rho \alpha } \nabla_{\alpha }A_2{}_{\mu \nu } \\\nonumber
			& + \bigl(c_1 -  c_1 \theta - 2 b_1 (-1 + \theta) \theta + c_2 (2 + \theta) + b_2 \theta (2 + \theta)\bigr) A_1{}_{\nu }{}^{\alpha } \nabla_{\alpha }A_2{}_{\mu }{}^{\rho } + (-3 c_3 + 3 b_3 \theta) A_1{}_{\mu }{}^{\rho } \nabla_{\alpha }A_2{}_{\nu }{}^{\alpha } \\\nonumber
			& + \bigl(c_1 (-1 + \theta) + 2 b_1 (-1 + \theta) \theta -  c_2 (2 + \theta) -  b_2 \theta (2 + \theta)\bigr) A_1{}_{\mu }{}^{\alpha } \nabla_{\alpha }A_2{}_{\nu }{}^{\rho } - 3 (2 c_3 + b_3 \theta) A_1{}_{\mu \nu } \nabla_{\alpha }A_2{}^{\rho \alpha } \\\nonumber
			& + \bigl(c_1 -  c_4 - 2 c_2 (-1 + \theta) + 2 c_1 \theta + c_4 \theta -  \theta (b_1 + b_2 + 2 b_1 \theta -  b_2 \theta)\bigr) A_2{}^{\rho \alpha } \nabla_{\mu }A_1{}_{\nu \alpha } \\\nonumber
			& + \bigl(c_2 + c_4 + c_1 (-1 + \theta) -  c_2 \theta -  c_4 \theta + \theta (b_1 + b_2 + 2 b_1 \theta -  b_2 \theta)\bigr) A_2{}_{\nu }{}^{\alpha } \nabla_{\mu }A_1{}^{\rho }{}_{\alpha } \\\nonumber
			& + \bigl(c_4 + 2 c_2 (-1 + \theta) -  c_4 \theta -  c_1 (1 + 2 \theta) + \theta (b_1 + b_2 + 2 b_1 \theta -  b_2 \theta)\bigr) A_1{}^{\rho \alpha } \nabla_{\mu }A_2{}_{\nu \alpha } \\\nonumber
			& + \bigl(c_1 -  c_4 + c_2 (-1 + \theta) -  c_1 \theta + c_4 \theta -  \theta (b_1 + b_2 + 2 b_1 \theta -  b_2 \theta)\bigr) A_1{}_{\nu }{}^{\alpha } \nabla_{\mu }A_2{}^{\rho }{}_{\alpha } \\\nonumber
			& + \bigl(c_4 + 2 c_2 (-1 + \theta) -  c_4 \theta -  c_1 (1 + 2 \theta) + \theta (b_1 + b_2 + 2 b_1 \theta -  b_2 \theta)\bigr) A_2{}^{\rho \alpha } \nabla_{\nu }A_1{}_{\mu \alpha } \\\nonumber
			& + \bigl(c_1 -  c_4 + c_2 (-1 + \theta) -  c_1 \theta + c_4 \theta -  \theta (b_1 + b_2 + 2 b_1 \theta -  b_2 \theta)\bigr) A_2{}_{\mu }{}^{\alpha } \nabla_{\nu }A_1{}^{\rho }{}_{\alpha } \\\nonumber
			& + \bigl(c_1 - c_4 - 2 c_2 (-1 + \theta) + 2 c_1 \theta + c_4 \theta -  \theta (b_1 + b_2 + 2 b_1 \theta -  b_2 \theta)\bigr) A_1{}^{\rho \alpha } \nabla_{\nu }A_2{}_{\mu \alpha } \\\nonumber
			& + \bigl(c_2 + c_4 + c_1 (-1 + \theta) -  c_2 \theta -  c_4 \theta + \theta (b_1 + b_2 + 2 b_1 \theta -  b_2 \theta)\bigr) A_1{}_{\mu }{}^{\alpha } \nabla_{\nu }A_2{}^{\rho }{}_{\alpha } \\\nonumber
			& + \bigl(c_2 (-1 + \theta) -  c_1 (2 + \theta) + c_4 (2 + \theta) -  \theta (b_1 + b_2 + 2 b_1 \theta -  b_2 \theta)\bigr) A_2{}_{\nu }{}^{\alpha } \nabla^{\rho }A_1{}_{\mu \alpha } \\\nonumber
			& + \bigl(c_2 -  c_2 \theta + c_1 (2 + \theta) -  c_4 (2 + \theta) + \theta (b_1 + b_2 + 2 b_1 \theta -  b_2 \theta)\bigr) A_2{}_{\mu }{}^{\alpha } \nabla^{\rho }A_1{}_{\nu \alpha } \\\nonumber
			& + \bigl(c_2 -  c_2 \theta + c_1 (2 + \theta) -  c_4 (2 + \theta) + \theta (b_1 + b_2 + 2 b_1 \theta -  b_2 \theta)\bigr) A_1{}_{\nu }{}^{\alpha } \nabla^{\rho }A_2{}_{\mu \alpha } \\\nonumber
			& + \bigl(c_2 (-1 + \theta) -  c_1 (2 + \theta) + c_4 (2 + \theta) -  \theta (b_1 + b_2 + 2 b_1 \theta - b_2 \theta)\bigr) A_1{}_{\mu }{}^{\alpha } \nabla^{\rho }A_2{}_{\nu \alpha } \, .
	\end{align}
}
By requiring this commutator to have the same functional form of Eq.\ \eqref{eq:ansatz-deltaAPi} we derive some relations for the dimensionless coefficients used to parametrize the most general transformation, which read
\begin{equation}
	\begin{split}
		& c_1 = b_{1} \, , \quad c_2 = -\frac{b_1}{3} \, , \quad c_3 = 0 \, ;\\
		& c_4 = b_1 \, , \quad b_2 = - \frac{2 b_1}{3}\, , \quad b_3 = 0 \, .
	\end{split}
\end{equation}
Finally, the angle is bound to satisfy
\begin{equation}
	 \theta= \frac{1}{4} \, .
\end{equation}

\section{Deriving the invariant action}\label{sect:app-derivation-action}

Let us consider the most general quartic Lagrangian for the torsion, which reads
\begin{equation}
	\begin{split}
		{\cal L}_{\rm quar} = \, & Q_{1}^{} T_{\alpha }{}^{\nu \sigma} T^{\alpha \lambda \mu} T_{\lambda \nu}{}^{\beta} T_{\mu \sigma \beta} + Q_{2}^{} T_{\alpha }{}^{\nu \sigma} T^{\alpha \lambda \mu} T_{\lambda \mu}{}^{\beta} T_{\nu \sigma \beta} + Q_{3}^{} T_{\alpha \lambda}{}^{\nu} T^{\alpha \lambda \mu} T_{\mu}{}^{\sigma \beta} T_{\nu \sigma \beta} \\
		& + Q_{4}^{} T^{\alpha \lambda \mu} T_{\lambda \alpha }{}^{\nu} T_{\mu}{}^{\sigma \beta} T_{\nu \sigma \beta} + Q_{5}^{} T^{\alpha }{}_{\alpha }{}^{\lambda} T_{\mu}{}^{\sigma \beta} T^{\mu}{}_{\lambda}{}^{\nu} T_{\nu \sigma \beta} + Q_{6}^{} T_{\alpha \lambda \mu} T^{\alpha \lambda \mu} T_{\nu \sigma \beta} T^{\nu \sigma \beta}\\
		& + Q_{7}^{} T^{\alpha }{}_{\alpha }{}^{\lambda} T^{\mu}{}_{\lambda \mu} T_{\nu \sigma \beta} T^{\nu \sigma \beta} + Q_{8}^{} T^{\alpha \lambda \mu} T_{\lambda}{}^{\nu \sigma} T_{\mu \nu}{}^{\beta} T_{\sigma \alpha \beta} + Q_{9}^{} T^{\alpha \lambda \mu} T_{\lambda \mu}{}^{\nu} T_{\nu}{}^{\sigma \beta} T_{\sigma \alpha \beta}\\
		& + Q_{10}^{} T_{\alpha }{}^{\nu \sigma} T^{\alpha \lambda \mu} T_{\lambda \nu}{}^{\beta} T_{\sigma \mu \beta} + Q_{11}^{} T^{\alpha }{}_{\alpha }{}^{\lambda} T^{\mu}{}_{\lambda}{}^{\nu} T_{\nu}{}^{\sigma \beta} T_{\sigma \mu \beta} + Q_{12}^{} T_{\alpha \lambda}{}^{\nu} T^{\alpha \lambda \mu} T_{\mu}{}^{\sigma \beta} T_{\sigma \nu \beta}\\
		& + Q_{13}^{} T^{\alpha \lambda \mu} T_{\lambda \alpha }{}^{\nu} T_{\mu}{}^{\sigma \beta} T_{\sigma \nu \beta} + Q_{14}^{} T^{\alpha }{}_{\alpha }{}^{\lambda} T_{\lambda}{}^{\mu \nu} T_{\mu}{}^{\sigma \beta} T_{\sigma \nu \beta} + Q_{15}^{} T^{\alpha }{}_{\alpha }{}^{\lambda} T_{\mu}{}^{\sigma \beta} T^{\mu}{}_{\lambda}{}^{\nu} T_{\sigma \nu \beta}\\
		& + Q_{16}^{} T_{\alpha \lambda \mu} T^{\alpha \lambda \mu} T^{\nu \sigma \beta} T_{\sigma \nu \beta} + Q_{17}^{} T^{\alpha \lambda \mu} T_{\lambda \alpha \mu} T^{\nu \sigma \beta} T_{\sigma \nu \beta} + Q_{18}^{} T^{\alpha }{}_{\alpha }{}^{\lambda} T^{\mu}{}_{\lambda \mu} T^{\nu \sigma \beta} T_{\sigma \nu \beta}\\
		& + Q_{19}^{} T_{\alpha \lambda}{}^{\nu} T^{\alpha \lambda \mu} T_{\sigma \nu \beta} T^{\sigma}{}_{\mu}{}^{\beta} + Q_{20}^{} T^{\alpha }{}_{\alpha }{}^{\lambda} T^{\mu}{}_{\lambda}{}^{\nu} T_{\sigma \nu \beta} T^{\sigma}{}_{\mu}{}^{\beta} + Q_{21}^{} T^{\alpha \lambda \mu} T_{\lambda}{}^{\nu \sigma} T_{\nu \mu}{}^{\beta} T_{\beta \alpha \sigma}\\
		& + Q_{22}^{} T_{\alpha }{}^{\nu \sigma} T^{\alpha \lambda \mu} T_{\lambda \nu}{}^{\beta} T_{\beta \mu \sigma} + Q_{23}^{} T_{\alpha }{}^{\nu \sigma} T^{\alpha \lambda \mu} T_{\lambda \mu}{}^{\beta} T_{\beta \nu \sigma} + Q_{24}^{} T_{\alpha \lambda}{}^{\nu} T^{\alpha \lambda \mu} T^{\sigma}{}_{\mu}{}^{\beta} T_{\beta \nu \sigma}\\
		& + Q_{25}^{} T^{\alpha \lambda \mu} T_{\lambda \alpha }{}^{\nu} T^{\sigma}{}_{\mu}{}^{\beta} T_{\beta \nu \sigma} + Q_{26}^{} T^{\alpha }{}_{\alpha }{}^{\lambda} T^{\mu}{}_{\lambda}{}^{\nu} T^{\sigma}{}_{\mu}{}^{\beta} T_{\beta \nu \sigma} + Q_{27}^{} T_{\alpha }{}^{\nu \sigma} T^{\alpha \lambda \mu} T_{\beta \nu \sigma} T^{\beta}{}_{\lambda \mu}\\
		& + Q_{28}^{} T_{\alpha }{}^{\nu \sigma} T^{\alpha \lambda \mu} T_{\beta \mu \sigma} T^{\beta}{}_{\lambda \nu} + Q_{29}^{} T^{\alpha }{}_{\alpha }{}^{\lambda} T_{\lambda}{}^{\mu \nu} T_{\mu \nu}{}^{\sigma} T^{\beta}{}_{\sigma \beta} + Q_{30}^{} T^{\alpha }{}_{\alpha }{}^{\lambda} T_{\mu \nu}{}^{\sigma} T^{\mu}{}_{\lambda}{}^{\nu} T^{\beta}{}_{\sigma \beta}\\
		& + Q_{31}^{} T^{\alpha }{}_{\alpha }{}^{\lambda} T^{\mu}{}_{\lambda}{}^{\nu} T_{\nu \mu}{}^{\sigma} T^{\beta}{}_{\sigma \beta} + Q_{32}^{} T^{\alpha }{}_{\alpha }{}^{\lambda} T^{\mu}{}_{\lambda \mu} T^{\nu}{}_{\nu}{}^{\sigma} T^{\beta}{}_{\sigma \beta} + Q_{33}^{} T^{\alpha }{}_{\alpha }{}^{\lambda} T_{\lambda}{}^{\mu \nu} T^{\sigma}{}_{\mu \nu} T^{\beta}{}_{\sigma \beta} \, .
	\end{split}
\end{equation}
Then, the homogeneous part of the symmetry transformation maps this Lagrangian into an expression which is still quartic in the torsion. Obviously, such an expression must vanish in order to have an actual gauge symmetry, and this requirement translates into the following set of equations for the coefficients that appear in the interacting Lagrangian
\begin{equation}
	\begin{split}
		& Q_{10} = -\frac{9 Q_1}{8} \, , \quad Q_{11} = 0 \, , \quad Q_{14} = 0 \, , \quad Q_{15} = 0 \, , \quad Q_{17} = -\frac{5 Q_{16}}{6} \, ;\\
		& Q_{19} = -\frac{19 Q_{12}}{12} - \frac{35 Q_{13}}{24} \, , \quad Q_2 = \frac{5 Q_{12}}{2} + \frac{7 Q_{13}}{2} \, , \quad Q_{20} = 0 \, ;\\
		& Q_{21} = -\frac{9 Q_1}{16} \, , \quad Q_{22} = \frac{11 Q_1}{4} \, , \quad Q_{23} = 7 Q_{12} + 9 Q_{13} \, ;\\
		& Q_{24} = \frac{7 Q_{12}}{3} + \frac{25 Q_{13}}{12} \, , \quad Q_{25} = -\frac{5 Q_{12}}{6} - \frac{17 Q_{13}}{24} \, , \quad Q_{26} = 0 \, ;\\
		& Q_{27} = \frac{23 Q_{12}}{12} + \frac{29 Q_{13}}{12} \, , \quad Q_{28} = \frac{9 Q_1}{16} \, , \quad Q_3 = \frac{7 Q_{12}}{6} + \frac{5 Q_{13}}{3} \, ;\\
		& Q_4 = -\frac{4 Q_{12}}{3} - \frac{11 Q_{13}}{6} \, , \quad Q_5 = 0 \, , \quad Q_6 = -\frac{3 Q_{16}}{10} \, , \quad Q_8 = -2 Q_1 \, ;\\
		& Q_9 = -\frac{5 Q_{12}}{2} - \frac{7 Q_{13}}{2}  \, , \quad Q_{11} = 0 \, , \quad Q_{14} = 0\, , \quad Q_{15} = 0 \, , \quad Q_{18} = 0 \, ;\\
		& Q_{20} = 0 \, , \quad Q_{26} = 0 \, , \quad Q_{29} = 0 \, , \quad Q_{30} = 0 \, , \quad Q_{31} = 0 \, , \quad Q_{32} = 0 \, ;\\
		& Q_{33} = 0 \, , \quad Q_5 = 0 \, , \quad Q_ 7 = 0  \, .
	\end{split}
\end{equation}
Thus, we have reduced the number of independent couplings in the fourth-order interacting Lagrangian from $33$ to $4$.

Let us now consider the cubic terms. We parametrize the most general Lagrangian cubic in the torsion as
\begin{equation}
	\begin{split}
		{\cal L}_{\rm cub} = & \, F_{1}^{} T^{\alpha \beta \mu} T_{\beta}{}^{\nu \rho} \nabla_{\alpha }T_{\mu \nu \rho} + F_{2}^{} T^{\alpha \beta \mu} T^{\nu}{}_{\beta}{}^{\rho} \nabla_{\alpha }T_{\mu \nu \rho} + F_{7}^{} T^{\alpha \beta \mu} T_{\beta}{}^{\nu \rho} \nabla_{\mu}T_{\alpha \nu \rho}\\
		& + F_{8}^{} T^{\alpha \beta \mu} T^{\nu}{}_{\beta}{}^{\rho} \nabla_{\mu}T_{\alpha \nu \rho} + F_{10}^{} T^{\alpha }{}_{\alpha }{}^{\beta} T^{\mu \nu \rho} \nabla_{\mu}T_{\beta \nu \rho} + F_{12}^{} T^{\alpha }{}_{\alpha }{}^{\beta} T^{\mu \nu \rho} \nabla_{\mu}T_{\nu \beta \rho}\\
		& + F_{13}^{} T^{\alpha }{}_{\alpha }{}^{\beta} T^{\mu}{}_{\beta}{}^{\nu} \nabla_{\mu}T^{\rho}{}_{\nu \rho} + F_{14}^{} T^{\alpha \beta \mu} T^{\nu}{}_{\beta}{}^{\rho} \nabla_{\nu}T_{\alpha \mu \rho} + F_{15}^{} T^{\alpha \beta \mu} T^{\nu}{}_{\beta}{}^{\rho} \nabla_{\nu}T_{\mu \alpha \rho} \\
		& + F_{22}^{} T^{\alpha }{}_{\alpha }{}^{\beta} T^{\mu}{}_{\beta}{}^{\nu} \nabla_{\nu}T^{\rho}{}_{\mu \rho} + F_{23}^{} T^{\alpha \beta \mu} T_{\beta}{}^{\nu \rho} \nabla_{\rho}T_{\alpha \mu \nu} + F_{25}^{} T^{\alpha \beta \mu} T_{\beta \mu}{}^{\nu} \nabla_{\rho}T_{\alpha \nu}{}^{\rho} \\
		& + F_{26}^{} T^{\alpha \beta \mu} T^{\nu}{}_{\beta \mu} \nabla_{\rho}T_{\alpha \nu}{}^{\rho} + F_{28}^{} T^{\alpha }{}_{\alpha }{}^{\beta} T^{\mu \nu \rho} \nabla_{\rho}T_{\beta \mu \nu} + F_{29}^{} T^{\alpha }{}_{\alpha }{}^{\beta} T^{\mu}{}_{\mu}{}^{\nu} \nabla_{\rho}T_{\beta \nu}{}^{\rho} \\
		& + F_{31}^{} T^{\alpha \beta \mu} T^{\nu}{}_{\beta}{}^{\rho} \nabla_{\rho}T_{\mu \alpha \nu} + F_{32}^{} T^{\alpha }{}_{\alpha }{}^{\beta} T^{\mu \nu \rho} \nabla_{\rho}T_{\mu \beta \nu} + F_{33}^{} T_{\alpha \beta}{}^{\nu} T^{\alpha \beta \mu} \nabla_{\rho}T_{\mu \nu}{}^{\rho} \\
		& + F_{34}^{} T^{\alpha \beta \mu} T_{\beta \alpha }{}^{\nu} \nabla_{\rho}T_{\mu \nu}{}^{\rho} + F_{35}^{} T^{\alpha }{}_{\alpha }{}^{\beta} T_{\beta}{}^{\mu \nu} \nabla_{\rho}T_{\mu \nu}{}^{\rho} + F_{36}^{} T^{\alpha }{}_{\alpha }{}^{\beta} T^{\mu}{}_{\beta}{}^{\nu} \nabla_{\rho}T_{\mu \nu}{}^{\rho} \\
		& + F_{37}^{} T^{\alpha \beta \mu} T_{\beta}{}^{\nu \rho} \nabla_{\rho}T_{\nu \alpha \mu} + F_{38}^{} T^{\alpha \beta \mu} T_{\beta \mu}{}^{\nu} \nabla_{\rho}T_{\nu \alpha }{}^{\rho} + F_{39}^{} T^{\alpha }{}_{\alpha }{}^{\beta} T^{\mu \nu \rho} \nabla_{\rho}T_{\nu \beta \mu} \\
		& + F_{40}^{} T^{\alpha }{}_{\alpha }{}^{\beta} T^{\mu}{}_{\beta}{}^{\nu} \nabla_{\rho}T_{\nu \mu}{}^{\rho} + F_{41}^{} T_{\alpha \beta \mu} T^{\alpha \beta \mu} \nabla_{\rho}T^{\nu}{}_{\nu}{}^{\rho} + F_{42}^{} T^{\alpha \beta \mu} T_{\beta \alpha \mu} \nabla_{\rho}T^{\nu}{}_{\nu}{}^{\rho} \\
		& + F_{43}^{} T^{\alpha }{}_{\alpha }{}^{\beta} T^{\mu}{}_{\beta \mu} \nabla_{\rho}T^{\nu}{}_{\nu}{}^{\rho} + F_{44}^{} T^{\alpha \beta \mu} T_{\beta \mu}{}^{\nu} \nabla_{\rho}T^{\rho}{}_{\alpha \nu} + F_{45}^{} T^{\alpha }{}_{\alpha }{}^{\beta} T_{\beta}{}^{\mu \nu} \nabla_{\rho}T^{\rho}{}_{\mu \nu} \\
		& + F_{46}^{} T^{\alpha }{}_{\alpha }{}^{\beta} T^{\mu}{}_{\beta}{}^{\nu} \nabla_{\rho}T^{\rho}{}_{\mu \nu} \, .
	\end{split}
\end{equation}
There are $31$ non-trivial contractions, and their coupling constant have subscripts which are not always consecutive because \texttt{All Contraction} is not able to identify boundary terms, which must be canceled out after generating the ansatz. By simultaneously applying a homogeneous transformation to the cubic terms and an affine one to the quartic ones we obtain tensor structure of the same type. Thus, we can constrain the coefficients of the previous equation by imposing invariance under this combined transformation of the two interacting Lagrangians. This procedure yields
\begin{equation}
	\begin{split}
		& F_1 = -\frac{F_{23}}{5} \, , \quad F_{10} = 0 \, , \quad F_{12} = 0 \, , \quad F_{14} = F_{23} \, , \quad F_{15} = -\frac{4 F_{23}}{5} \, ;\\
		& F_2 = \frac{6 F_{23}}{5} \, , \quad F_{25} = \frac{3 F_{23}}{10} \, , \quad F_{26} = \frac{F_{23}}{10} \, , \quad F_{28} = 0 \, , \quad F_{31} = \frac{4 F_{23}}{5} \, ;\\
		& F_{32} = 0 \, , \quad F_{33} = -\frac{F_{23}}{2}  \, , \quad F_{34} = -\frac{2 F_{23}}{5} \, , \quad F_{35} = 0 \, , \quad F_{36} = 0 \, , \quad F_{37} = \frac{3 F_{23}}{5} \, ;\\
		& F_{38} = -\frac{3 F_{23}}{10} \, , \quad F_{39} = 0 \, , \quad F_{40} = 0 \, , \quad F_{41} = 0 \, , \quad F_{42} = 0 \, , \quad F_{44} = \frac{3 F_{23}}{5} \, ;\\
		& F_{45} = 0 \, , \quad F_{46} = 0 \, , \quad F_7 = \frac{F_{23}}{5} \, , \quad F_8 = -\frac{3 F_{23}}{5} \, , \quad Q_1 = -\frac{4 \zeta F_{23}}{15} \, ;\\
		& Q_{12} = -\frac{\zeta F_{23}}{2} \, , \quad Q_{13} = \frac{2 \zeta F_{23}}{5} \, , \quad Q_{16} = 0  \, , \quad F_{10} = 0 \, , \quad F_{12} = 0 \, ;\\
		& F_{13} = 0 \, , \quad F_{22} = 0 \, , \quad F_{29} = 0 \, , \quad F_{43} = 0 \, .
	\end{split}
\end{equation}
This result not only tells us that there is only one independent coupling ($F_{23}$) in the cubic Lagrangian, but it also fixes the four independent couplings of the quartic Lagrangian in terms of $F_{23}$ and the charge $\zeta$.

Now we take into account the quadratic terms in the torsion, which comprise kinetic terms, mass terms and the interaction terms with the Riemann tensor and its derivatives. This part of the Lagrangian can be written as
\begin{equation}
	\begin{split}
		{\cal L}_{\rm quad} = & \, C_{1} T_{\alpha \beta \mu} T^{\alpha \beta \mu} + B_{1} R T_{\alpha \beta \mu} T^{\alpha \beta \mu} + B_{2} R_{\beta \mu \nu \rho} T_{\alpha }{}^{\nu \rho} T^{\alpha \beta \mu} \\
		& + C_{2} T^{\alpha \beta \mu} T_{\beta \alpha \mu} + B_{4} R T^{\alpha \beta \mu} T_{\beta \alpha \mu} + B_{5} R^{\alpha \beta} T_{\alpha }{}^{\mu \nu} T_{\beta \mu \nu} + B_{6} R_{\alpha \mu \nu \rho} T^{\alpha \beta \mu} T_{\beta}{}^{\nu \rho} \\
		& + B_{8} R^{\alpha \beta} T_{\alpha }{}^{\mu \nu} T_{\mu \beta \nu} + B_{9} R^{\alpha \beta} T_{\mu \beta \nu} T^{\mu}{}_{\alpha }{}^{\nu} + C_{3} T^{\alpha }{}_{\alpha }{}^{\beta} T^{\mu}{}_{\beta \mu} + B_{10} R T^{\alpha }{}_{\alpha }{}^{\beta} T^{\mu}{}_{\beta \mu} \\
		& + B_{11} R_{\beta \mu \nu \rho} T^{\alpha }{}_{\alpha }{}^{\beta} T^{\mu \nu \rho} + B_{13} R^{\alpha \beta} T^{\mu}{}_{\alpha }{}^{\nu} T_{\nu \beta \mu} + B_{14} R^{\alpha \beta} T^{\mu}{}_{\alpha \mu} T^{\nu}{}_{\beta \nu}\\
		& + B_{15} R_{\alpha \mu \nu \rho} T^{\alpha \beta \mu} T^{\nu}{}_{\beta}{}^{\rho} + B_{16} R_{\alpha \nu \mu \rho} T^{\alpha \beta \mu} T^{\nu}{}_{\beta}{}^{\rho} + B_{18} R^{\alpha \beta} T_{\alpha \beta}{}^{\mu} T^{\nu}{}_{\mu \nu} \\
		&  -  C_{12} \nabla_{\mu}T^{\nu}{}_{\beta \nu} \nabla^{\mu}T^{\alpha }{}_{\alpha }{}^{\beta} -  C_8 \nabla_{\beta}T^{\alpha \beta \mu} \nabla_{\nu}T_{\alpha \mu}{}^{\nu} -  C_7 \nabla_{\alpha }T^{\alpha \beta \mu} \nabla_{\nu}T_{\beta \mu}{}^{\nu} \\
		&  -  C_9 \nabla_{\beta}T^{\alpha \beta \mu} \nabla_{\nu}T_{\mu \alpha }{}^{\nu} + C_{11} \nabla_{\beta}T^{\alpha }{}_{\alpha }{}^{\beta} \nabla_{\nu}T^{\mu}{}_{\mu}{}^{\nu} + C_6 \nabla_{\alpha }T^{\alpha \beta \mu} \nabla_{\nu}T^{\nu}{}_{\beta \mu} \\
		& -  C_{10} \nabla^{\mu}T^{\alpha }{}_{\alpha }{}^{\beta} \nabla_{\nu}T^{\nu}{}_{\beta \mu} + C_4 \nabla_{\nu}T_{\alpha \beta \mu} \nabla^{\nu}T^{\alpha \beta \mu} + C_5 \nabla_{\nu}T_{\beta \alpha \mu} \nabla^{\nu}T^{\alpha \beta \mu} \, .
	\end{split}
\end{equation}
Once again we use the strategy of applying the homogeneous transformation to the quadratic terms and the affine one to cubic interactions, and imposing the invariance of the resulting expression we obtain the following relations for the $B$'s
\begin{equation}
	\begin{split}
		& B_2=\frac{B_{15}}{8} \, , \quad B_4 = -\frac{5 B_1}{3} \, , \quad B_5 = -2 B_{13} \, , \quad B_6=-\frac{3 B_{15}}{4} \, , \quad  B_8 = 4 B_{13} \, ;\\
		& B_9 = -B_{13} \, , \quad B_{10} = 0 \, , \quad B_{14} = 0 \, , \quad B_{16} = \frac{B_{15}}{4} \, , \quad B_{18} = 0 \, . 
	\end{split}
\end{equation}
On the other hand, we derive the subsequent equations for the $C$'s, i.e.,
\begin{equation}
	\begin{split}
		& C_2 = -\frac{5 C_1}{3} \, , \quad \quad C_3 = 0 \, , \quad C_4 = -\frac{3 B_{15}}{8} \, , \quad C_5 = \frac{5 B_{15}}{8} \, , \quad C_6 = \frac{B_{15}}{4} \, ;\\
		& C_7 = - \frac{B_{15}}{2}  \, , \quad  C_8 = \frac{B_{15}}{8} \, , \quad C_9 = -\frac{B_{15}}{8} \, , \quad C_{10} = 0 \, , \quad C_{11} = 0 \, , \quad C_{12} = 0 \, .
	\end{split}
\end{equation}
Finally, the $F_{23}$ coupling is written in terms of $B_{15}$ and the charge as
\begin{equation}
	 F_{23} = -\frac{5 \zeta B_{15}}{6} \, .
\end{equation}

The last terms that we must consider are those linear in the torsion. After repeatedly using the differential Bianchi identities one can see that the most general parametrization of these terms displays only two independent tensor structures, i.e.,
\begin{equation}
	{\cal L}_{\rm lin} = D_1 T^\nu{}_{\mu\nu} \nabla^\mu R - 2 D_2 T^{\rho\mu\nu} \nabla_\nu R_{\mu\rho} \, .
\end{equation}
In analogy with what we have done in the previous steps we equate the affine variation of the quadratic terms and the homogeneous variation of the linear ones, and we impose that the resulting tensor structure vanishes. By doing so we find
\begin{equation}
	B_1=0 \, , \quad B_{13} = \frac{\zeta \, D_2}{3} \, , \quad B_{15} = \frac{8 \zeta \, D_2}{3} \, \quad C_1 = 0 \, , \quad D_2=0 \, .
\end{equation}
For notational convenience we have chosen to set $D_2=\xi$ in the main text. Thus, by rescaling the torsion to normalize the kinetic terms we deduce that the action is written only in terms of the ratio $\xi/\zeta$.


\bibliographystyle{chetref}

\end{document}